\documentclass[printer]{aa}  
\usepackage{graphicx}
\usepackage{txfonts}
\usepackage{enumitem}
\usepackage[breaklinks, colorlinks, citecolor=blue]{hyperref}
\usepackage{multirow}
\usepackage{natbib}

\usepackage{color, soul}
\setcitestyle{citesep={,}}

\usepackage[normalem]{ulem}

\begin{document}

   \title{LOFAR Deep Fields: Probing faint Galactic \\ polarised emission in ELAIS-N1 \thanks{The Faraday cube is only available in electronic form at the CDS via anonymous ftp to \url{cdsarc.u-strasbg.fr}.}}
   \subtitle{}
   \titlerunning{Probing faint Galactic polarised emission in ELAIS-N1}
   \authorrunning{I. Šnidarić et al.}

   \author{Iva \v{S}nidari\'c\inst{\ref{inst1}\dagger}
          \and
          Vibor Jeli\'c\inst{\ref{inst1}\ddagger}
          \and
          Maaijke Mevius\inst{\ref{inst2}}
          \and
          Michiel Brentjens\inst{\ref{inst2}}
          \and
          Ana Erceg\inst{\ref{inst1}}
          \and
          Timothy W. Shimwell\inst{\ref{inst2},\ref{inst3}}
          \and
          Sara Piras\inst{\ref{inst4}}
          \and
          Cathy Horellou\inst{\ref{inst4}}
          \and
          Jose Sabater\inst{\ref{inst5}}
          \and
          Philip N. Best\inst{\ref{inst5}}
          \and
          Andrea Bracco\inst{\ref{inst6}}
          \and
          Lana Ceraj\inst{\ref{inst1}}
          \and
          Marijke Haverkorn\inst{\ref{inst7}}
          \and
          Shane P. O'Sullivan\inst{\ref{inst8}}
          \and
          Luka Turi\'c\inst{\ref{inst1}}
           \and
          Valentina Vacca\inst{\ref{inst9}}
          }

   \institute{Ru{\dj}er Bo\v{s}kovi\'c Institute, Bijeni\v{c}ka cesta 54, 10000 Zagreb, Croatia \\
              \email{$^\dagger$isnidar@irb.hr, $^\ddagger$vibor@irb.hr} \label{inst1} \and
              ASTRON, Netherlands Institute for Radio Astronomy, Oude Hoogeveensedijk 4, 7991 PD, Dwingeloo, The Netherlands \label{inst2} \and
              Leiden Observatory, Leiden University, P.O. Box 9513, 2300 RA Leiden, The Netherlands \label{inst3}\and
              Department of Space, Earth and Environment, Chalmers University of Technology, Onsala Space Observatory, 43992 Onsala, Sweden
               \label{inst4}\and
              Institute for Astronomy, University of Edinburgh, Royal Observatory, Blackford Hill, Edinburgh, EH9 3HJ, UK \label{inst5}\and
              Laboratoire de Physique de l'Ecole Normale Sup\'erieure, ENS, Universit\'e PSL, CNRS, Sorbonne Universit\'e, Universit\'e de Paris, F-75005 Paris, France \label{inst6}\and
              Department of Astrophysics/IMAPP, Radboud University, P.O. Box 9010, 6500 GL Nijmegen, The Netherlands\label{inst7} \and
              School of Physical Sciences and Centre for Astrophysics \& Relativity, Dublin City University, Glasnevin, D09 W6Y4, Ireland \label{inst8}\and
              INAF-Osservatorio Astronomico di Cagliari, Via della Scienza 5, I-09047 Selargius (CA), Italy \label{inst9}
             }
   \date{Received XXX; accepted XXX}

  \abstract
 {We present the first deep polarimetric study of Galactic synchrotron emission at low radio frequencies. Our study is based on 21 observations of the European Large Area Infrared Space Observatory Survey-North 1 (ELAIS-N1) field using the Low-Frequency Array (LOFAR) at frequencies from 114.9 to 177.4 MHz. These data are a part of the LOFAR Two-metre Sky Survey Deep Fields Data Release 1. We used very low-resolution ($4.3'$) Stokes QU data cubes of this release. We applied rotation measure (RM) synthesis to decompose the distribution of polarised structures in Faraday depth, and cross-correlation RM synthesis to align different observations in Faraday depth. We stacked images of about 150 hours of the ELAIS-N1 observations to produce the deepest Faraday cube at low radio frequencies to date, tailored to studies of Galactic synchrotron emission and the intervening magneto-ionic interstellar medium.  This Faraday cube covers $\sim36~{\rm deg^{2}}$ of the sky and has a noise of $27~{\rm \mu Jy~PSF^{-1}~RMSF^{-1}}$ in polarised intensity. This is an improvement in noise by a factor of approximately the square root of the number of stacked data cubes ($\sim\sqrt{20}$), as expected, compared to the one in a single data cube based on five-to-eight-hour observations.  We detect a faint component of diffuse polarised emission in the stacked cube, which was not detected previously. Additionally, we verify the reliability of the ionospheric Faraday rotation corrections estimated from the satellite-based total electron content measurements to be of $~\sim0.05~{\rm rad~m^{-2}}$. We also demonstrate that diffuse polarised emission itself can be used to account for the relative ionospheric Faraday rotation corrections with respect to a reference observation.}


  \maketitle
%
 
\section{Introduction}\label{sec:intro}
The LOw-Frequency ARray \citep[LOFAR,][]{vanhaarlem13} is a radio interferometer utilising a new-generation phased-array design to explore the low-frequency radio sky ($10-240~{\rm MHz}$) in the Northern Hemisphere. This is carried out through a number of key science projects, including a series of ongoing LOFAR surveys. The LOFAR Two-meter Sky Survey \citep[LoTSS,][]{shimwell17,shimwell19,shimwell22} and LOFAR Low Band Antenna Sky Survey \citep[LoLSS,][]{deGasperin21} are the wide-area surveys at 120 -- 168 MHz and 42 -- 66 MHz, respectively. These surveys are complemented by a few deeper fields, known as the LoTSS-Deep Fields \citep{tasse21,sabater21} and the LoLSS-Deep Fields \citep{deGasperin21}. 

The deep fields are selected in regions covered by a wealth of multi-wavelength data and the first data release includes the Bo\"otes, Lockman Hole, and European Large Area Infrared Space Observatory Survey-North 1 (ELAIS-N1) fields. These data sets make it possible to probe a new, fainter population of radio sources dominated by star-forming galaxies and radio-quiet active galactic nuclei \citep{smolcic17, novak18, kondapally21}. In addition, they enable studies of the under-explored polarised radio source population at sub-milijansky flux densities at low frequencies. 

\citet{ruiz21} did an initial analysis of polarised radio sources in the LoTSS Deep Fields using six eight-hour observations of the ELAIS-N1 field. They used low-resolution ($20''$) polarimetric images \citep{sabater21} and successfully demonstrated the feasibility of stacking LOFAR data. While they detected three polarised sources in a single observation, this number increased by more than a factor of three for the stacked data; they reported detection of seven additional sources, yielding a surface density of polarised sources of one per 1.6 deg$^2$. This work is being extended through stacking of 19 eight-hour LoTSS Deep Field observations of ELAIS-N1 re-imaged at higher angular resolution ($6''$) to further decrease the detection threshold and increase the number of detected polarised sources and probe the polarised source counts in the sub-mJy regime (Piras et al., in prep.).

When stacking, polarisation data need to be first properly calibrated and corrected for the Faraday rotation in the Earth's ionosphere \citep{murray54,hatanaka56}. Ionospheric Faraday rotation is a time- and direction-dependent propagation effect proportional to the integral along the line of sight (LOS) of the product of the total electron content (TEC) of plasma in the ionosphere and a projection of the geomagnetic field, $\vec{B_\mathrm{geo}}$, to the LOS towards the observed field of view (FoV). It is characterised by the ionospheric rotation measure ($RM_\mathrm{ion}$), which, in the thin-shell model, can be approximated as \citep[e.g.][]{sotomayor13} 
\begin{equation}
\frac{RM_\mathrm{ion}}{\mathrm{[rad \ m^{-2}]}}=0.26 \frac{\mathrm{TEC_{LOS}}}{\mathrm{[TECU]}} \frac{B_\mathrm{geo,LOS}}{\mathrm{[G]}},
\end{equation}
where $\mathrm{TEC_{LOS}}$ is the total electron content, measured in TEC units (1 TECU = $10^{16}$ electrons m$^{-2}$), at the ionospheric piercing point of the LOS. A typical $RM_\mathrm{ion}$ is $0.5-2~{\rm rad~m^{-2}}$ \citep{sotomayor13, jelic14, jelic15} at moderate geographical latitudes during nighttime. Daytime values are higher due to solar irradiation and an increase in TEC. The TEC decreases after the sunset due to recombination of plasma in the ionosphere.

Given that ionospheric Faraday rotation changes the polarisation angle $\theta$ of the observed emission on timescales smaller than the total integration time of observation, the observed polarised emission may be incoherently added during the synthesis, resulting in partial, or, in exceptional cases, full depolarisation. Ionospheric depolarisation effects are mostly relevant at lower radio frequencies, as Faraday rotation is inversely proportional to a square of the frequency ($\Delta\theta \sim RM_\mathrm{ion}\nu^{-2}$). At 150 MHz, a change in the ionospheric Faraday rotation of $\sim 0.8~{\rm rad~m^{-2}}$ results in a $180^\circ$ rotation of the polarisation vector and therefore full depolarisation.

The LOFAR observations are usually corrected for the ionospheric Faraday rotation in a direction-independent manner by combining global geomagnetic field models with Global Navigation Satellite System (GNSS) observations of the ionospheric TEC \citep{sotomayor13, mevius2018}. This was first tested on the LOFAR commissioning observations of the ELAIS-N1 field \citep{jelic14}, and since then, it is widely used in polarisation studies with LOFAR \citep[e.g.][]{jelic15, vaneck17, turic21, erceg22}. Depending on the source of the TEC data, the estimated uncertainty in the calculated ionospheric Faraday rotation is within a factor of a few of $0.1~{\rm rad~m^{-2}}$ at time intervals of 15 minutes to two hours.

Recently, \citet{degasperin18} showed that LOFAR Low Band Antenna (LBA) station-based gain phase can be decomposed into a few systematic effects related to clock delays and ionospheric effects and used directly to obtain independent measurements of the absolute TEC.  The LOFAR measured TEC values are within $10\%$ of the satellite-based measurements and have two orders of magnitude better time resolution. This has enabled a new, efficient, unified calibration strategy for LOFAR LBA \citep{degasperin19}. However, further detailed analysis of systematic uncertainties related to ionospheric Faraday rotation corrections is needed, as well as, an assessment of the method for LOFAR High Band Antenna (HBA) observations and direction-dependent effects. 

The six ELAIS-N1 observations analysed by \citet{ruiz21} were corrected for the ionospheric Faraday rotation by the satellite-based TEC measurements \citep{sabater21}. To check how well they were corrected relatively to each other, \citet{ruiz21} compared the observed Faraday depth of the bright reference source in each observation and found a relative difference varying from $-0.12$ to $+0.05~{\rm rad~m^{-2}}$. Then they calculated the difference in the observed polarisation angle, corrected each observation accordingly, and stacked the data. 

A complementary method to check for a relative alignment between the observations concerning the Faraday rotation in the ionosphere is based on using the polarised diffuse Galactic synchrotron emission \citep{lenc16,brentjens18}. This type of emission is ubiquitous at low radio frequencies \citep[e.g.][and references therein]{erceg22} and allows analysis over a larger portion of the FoV compared to using a single reference polarised source. Ionospheric Faraday rotation corrections obtained in such a way should improve the accuracy of corrections and allow the analysis of differential variations across the field. 

In this work, we used the polarised diffuse synchrotron emission to study the ionospheric Faraday rotation corrections in 21 LOFAR observations of the ELAIS-N1 field. We also stacked very low-resolution images ($4.3'$) to study the faint component of the diffuse polarised emission in the ELAIS-N1 field, whose bright component was observed in the commissioning phase of the LOFAR \citep{jelic14}. The paper is organised as follows. LOFAR observations and related data products are described in Sect. 2. Section 3 presents the analysis of the ionospheric Faraday rotation corrections. Section 4 describes methodology for stacking the very low-resolution images. The final stacked Faraday cube is presented and analysed in Sect. 5. The newly detected faint polarised emission is discussed in Sect. 6. Summary and conclusions are presented in Sect. 7.

\section{Data and processing}
In this section, we describe the LoTSS-Deep Fields observations and the derived data products used in this paper. We also give an overview of the rotation measure (RM) synthesis technique and its parameters used to create Faraday cubes. 

\subsection{LoTSS-Deep Fields observations and very low-resolution images}
\label{sub:deep_filed_observations}
The ELAIS-N1 data analysed in this paper are part of the LoTSS-Deep Fields Data Release 1 \citep{sabater21}. We used 21 out of 27 observations, which were of good quality \citep[10 observations from Cycle 2 and 11 observations from Cycle 4, IDs 009--018, 020--024, 026--028, 030--032 in table 1 in][]{sabater21}. The data were taken with the LOFAR HBA from May 2014 to August 2015 (under project codes LC2$\_$024 and LC4$\_$008), covering the frequency range from 114.9 to 177.4 MHz dived into 320 frequency sub-bands. The observing time of each observation was between five and eight hours, taken during night-time and symmetric around transit.  The array was used in the HBA DUAL INNER configuration \citep{vanhaarlem13}. The HBA antennas of each core station are clustered in two groups of 24 tiles of 16 dual-polarised antennas. Each cluster of 24 tiles was then treated as an independent HBA core station. The remote stations have one group of 48 HBA tiles of 16 dual-polarised antennas. They were reduced to inner 24 tiles, to have the same shape and number of tiles as the dual core HBA stations. This provided a uniform general shape of the primary beam over the entirety of the LOFAR stations in the Netherlands. The phase centre of the main target field was at RA $16^{\rm h}11^{\rm m}00^{\rm s}$ and Dec $+55^{\circ}00'00''$ (J2000). 

Cycle 2 data were taken and pre-processed jointly with the LOFAR Epoch of Reionisation Key Science Project team in a slightly different way than Cycle 4 data. 
This created a difference in frequency configurations of the final data products of the two cycles. Here we give a brief overview of the main processing steps and relevant differences for each cycle, while details are provided in \citet{sabater21}.

The pre-processing of the data included averaging in time and frequency. Before averaging, the Cycle 2 data were automatically flagged for radio-frequency interference (RFI) using \texttt{AOFlagger} \citep{offringa12}. The first two and the last two frequency channels were then removed from each 64-channel sub-band to minimise the band-pass effects. The remaining 60 channels were averaged to 15 channels per sub-band. The Cycle 4 data were originally averaged by the observatory to 16 channels per sub-band, without discarding the channels at edges of each sub-band. After that, they were flagged for the RFI. The data from both cycles were averaged in time to $2~{\rm s}$. 

The direction-independent calibration was done using the \texttt{PREFACTOR pipeline} \citep{vanweeren16,degasperin19}, which corrects for the polarisation phase offset introduced by the station calibration table, the instrumental time delay associated with clocks in the remote stations, the amplitude band-pass, and ionospheric direction-independent delays and Faraday rotation. 

\begin{figure}[t!]
  \centering
  \includegraphics[width=\linewidth]{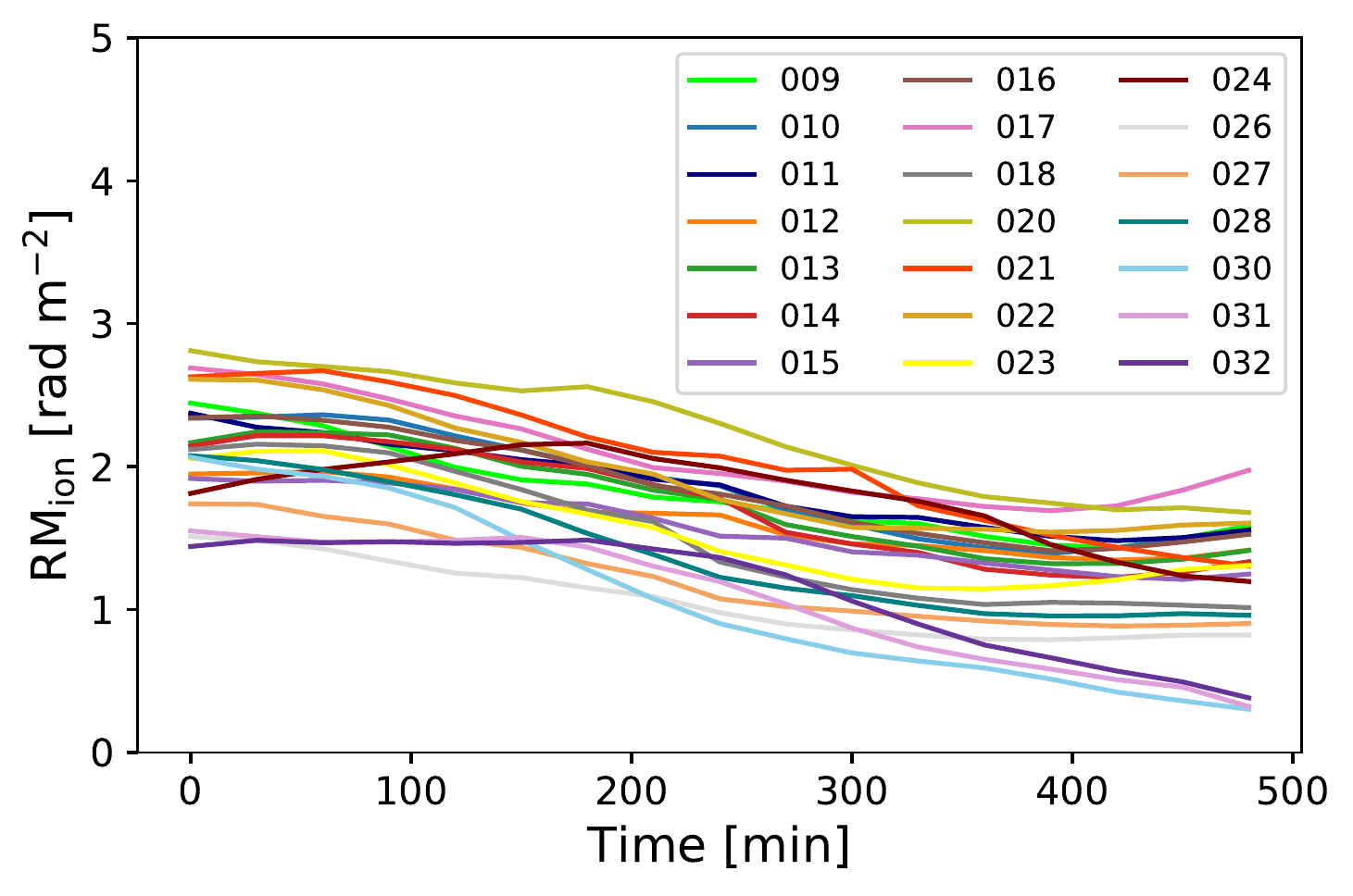}
  \caption{Calculated $RM_\mathrm{ion}$ corrections given at 30-minute intervals for different observations using the satellite-based TEC measurements and the global geomagnetic field model. The observed decrease of $RM_{\rm ion}$ during each nighttime observation is due to recombination of plasma in the ionosphere, which happens after the sunset and decreases the TEC throughout night.}
   \label{fig:RM_ion}
\end{figure}

The ionospheric Faraday rotation corrections were done by \texttt{RMextract} \citep{mevius2018}, which combines the satellite-based TEC measurements and the global geomagnetic field models to predict the corrections. 
The ionosphere was modelled as a single-phase screen above the array, taking into account its spatial structure. The $RM_\mathrm{ion}$ corrections were calculated for each LOFAR station separately; however, the model does not allow for direction-dependent corrections within the FoV. Moreover, the model uses a thin-screen approximation. The contribution from the plasmasphere to the total integrated electron content along the LOS can be significant \citep[up to $40\%$ at moderate geographical latitudes,][]{yizengaw2008}. Since the GNSS data include the full integrated electron density, including the plasmaspheric contribution, and the magnetic field contribution from the higher layers is significantly smaller, the derived RM values using a thin screen model are likely an overestimate of the ionospheric Faraday rotation. Figure~\ref{fig:RM_ion} shows calculated values for the LOFAR station CS002, as an example. The curves are given for different nights as a function of the observing time at 30-minute intervals. The absolute $RM_\mathrm{ion}$ values are between 0.5 and 3 ${\rm rad~m^{-2}}$, while their relative variations during observations are on average $0.9\pm0.3~{\rm rad~m^{-2}}$. 

The final step of processing included the direction-dependent calibration done by the \texttt{DDF pipeline} \citep{tasse21} and imaging of the data in full Stokes parameters (I, Q, U, and V). In this work, we used very low (vlow) resolution ($4.3'$) Stokes QU data cubes \citep{sabater21}. They are split in 800 or 640 frequencies of $73.24~{\rm kHz}$ or $97.66~{\rm kHz}$ width in the case of Cycle 2 or 4 data, respectively, due to their different frequency configurations. 
\begin{figure}[!t]
  \centering
  \includegraphics[width=0.5\textwidth]{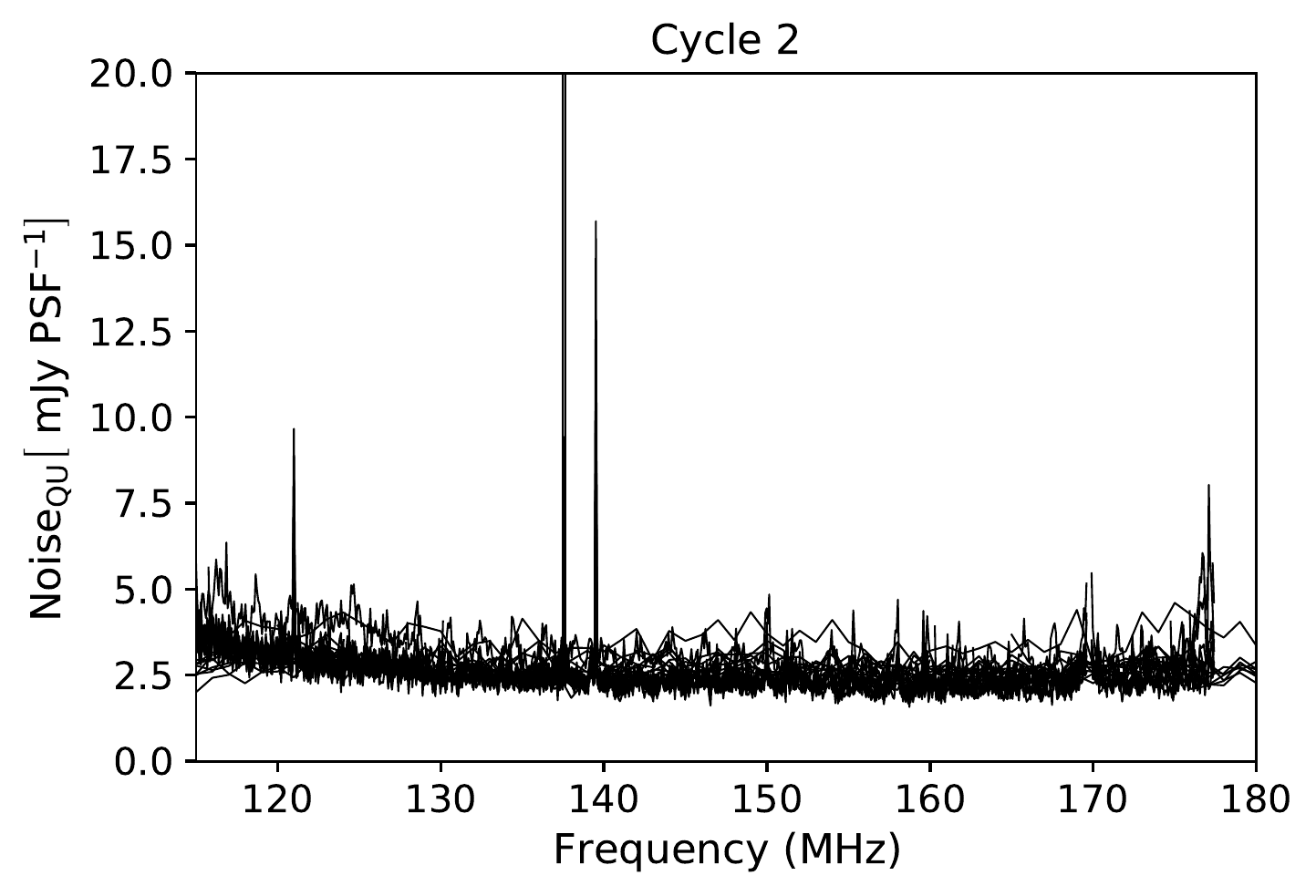}
  \includegraphics[width=0.5\textwidth]{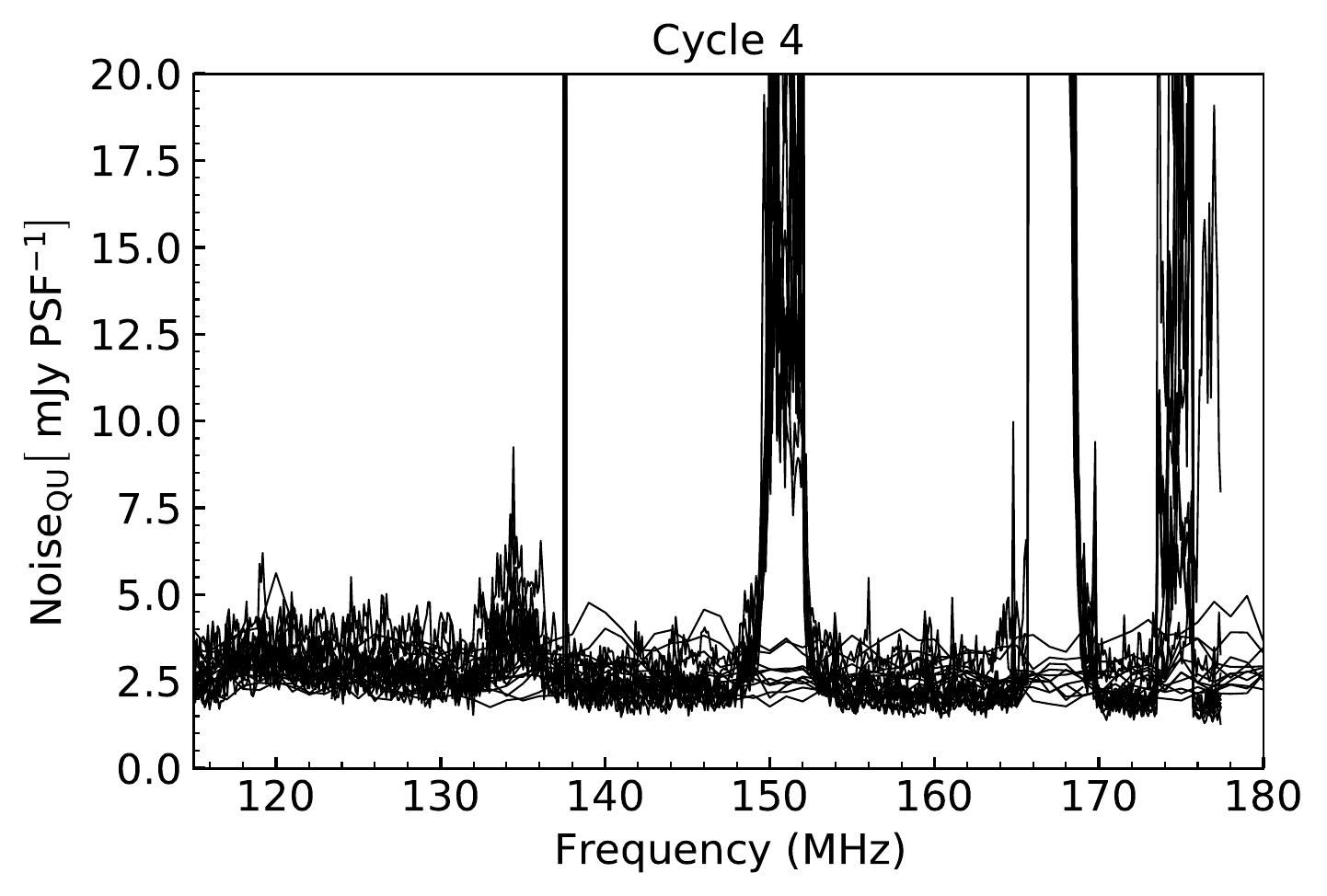}
  \caption{Noise in Stokes QU data cubes for observations from Cycle 2 (top plot) and Cycle 4 (bottom plot) as a function of frequency. Cycle 4 data are much more affected by broad RFI than Cycle 2 data due to DABs and DVBs.}
   \label{fig:noise_study}
\end{figure}

Figure~\ref{fig:noise_study} shows the noise in Stokes QU data cubes as a function of frequency. The noise at each frequency was calculated as a standard deviation in the corner of the primary-beam-uncorrected image (farthest out of the primary beam), where no polarised emission is present. The noise in Stokes Q and U is comparable. Cycle 4 data are much more affected by broad RFI \citep{offringa13} at frequencies around 134, 151, 167, and 174 MHz than Cycle 2 data. These RFIs are due to man-made wireless applications such as digital audio or video broadcasts (DABs or DVBs). Over the observed frequency range, a typical noise at frequencies not affected by the RFI is $\sim3.3~{\rm mJy~PSF^{-1}}$ in Cycle 2 data and $\sim2.7~{\rm mJy~PSF^{-1}}$ in Cycle 4 data. A small difference between the two cycles arises from their different frequency configurations and hence frequency channel widths.

\subsection{RM synthesis and Faraday depth cubes} \label{RMsynth}
We created Faraday data cubes of the ELAIS-N1 deep field observations for our analysis. They were produced by applying the RM synthesis technique \citep{burn66,brentjens05} to Stokes QU frequency data cubes. This technique decomposes the observed polarised emission by the amount of Faraday rotation of its polarisation angle, $\theta$, experienced at wavelength $\lambda$:
\begin{equation}\label{eq:FR}
     \frac{\Delta\theta}{\mathrm{[rad]}} = \frac{\Phi}{\mathrm{[rad~m^{-2}]}}\frac{\lambda^2}{\mathrm{[m^2]}}.
\end{equation}
The quantity $\Phi$ is called Faraday depth, and it is defined as
\begin{equation}\label{eq:FD}
     \frac{\Phi}{\mathrm{[rad \ m^{-2}]}} = 0.81 \int_{0}^{d} \frac{n_e}{\rm [cm^{-3}]} \frac{B_\parallel}{\mathrm{[\mu G]}} \frac{dl}{\mathrm{[pc]}},
\end{equation}
where $n_{e}$ is the density of thermal electrons and $B_\parallel$ is the magnetic field component parallel to the LOS. The integral is taken over the path length $dl$ from the source ($l=0$) to the observer ($l=d$). If the magnetic field component is pointing towards the observer, the value of the Faraday depth is positive and vice versa. Equation~\ref{eq:FD} and the sign convention related to the magnetic field component along the line of sight are in agreement with the correct sense of Faraday rotation discussed by \citet{ferriere21}.

For a given location in the sky, the RM synthesis gives us the distribution of the observed polarised emission in Faraday depth. This so-called Faraday spectrum is the Fourier transform of the complex polarisation of the observed signal, $P(\lambda^{2})=Q(\lambda^{2})+{\rm i}U(\lambda^{2})$, from $\lambda^2$- to $\Phi$-space \citep{brentjens05}: 
\begin{equation}\label{eq:CP}
     F(\Phi) = \frac{1}{W(\lambda^{2})}
     \int_{-\infty}^{+\infty} P(\lambda^{2})P^\ast(\lambda^{2})\exp^{-i2\Phi \lambda^{2}} d\lambda^2,
\end{equation}
where $W(\lambda^{2})$ is the non-zero-weighting function, usually taken to be 1 at $\lambda^2$ where measurements are taken and 0 elsewhere. If the RM synthesis is applied over a sky area, we can study the morphology of the observed polarised emission at different Faraday depths, to perform the so-called Faraday tomography. Characteristics of the $\lambda^{2}$ distribution constrain scales in Faraday depth that we can probe when performing the RM synthesis. A resolution in Faraday depth is inversely proportional to the spectral bandwidth ($\Delta\lambda^2$) as $\delta\Phi\approx2\sqrt{3}/\Delta\lambda^2$ and corresponds to the width of the rotation measure spread function \citep[RMSF,][]{brentjens05}. The maximum detectable Faraday scale is inversely proportional to the smallest ($\lambda_{\rm min}^2$) measured $\lambda^2$ as $\Delta\Phi_{\rm scale}\approx\pi/\lambda_{\rm min}^2$.

\begin{figure}
   \centering
   \includegraphics[width=\hsize]{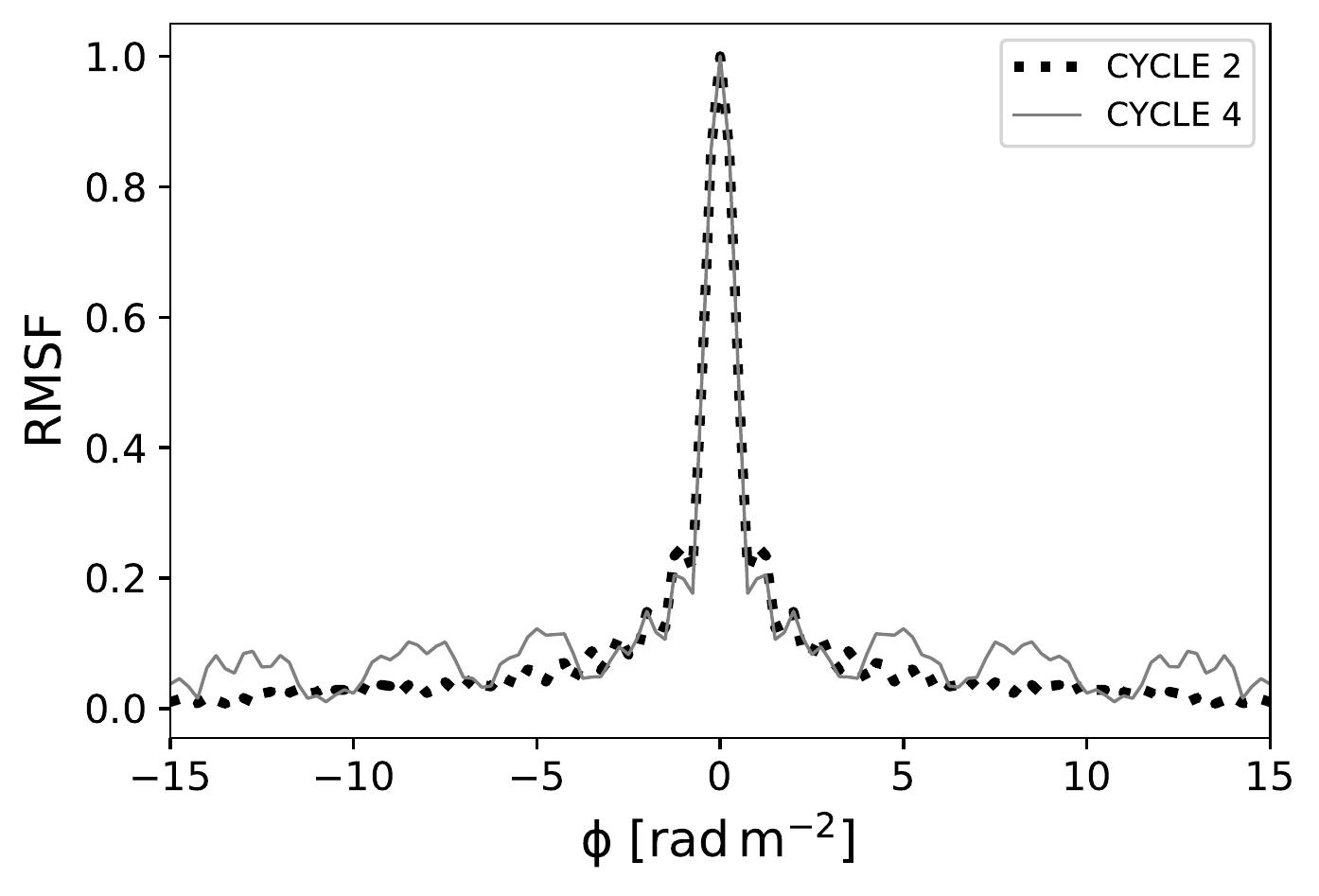}
      \caption{RMSF for ten observations in Cycle 2 (dashed) and 11 observations in Cycle 4 (solid line) of the ELAIS-N1 field.}
    \label{fig:rmsf}
\end{figure}

We used the publicly available code  \texttt{rm-synthesis}\footnote{\url{https://github.com/brentjens/rm-synthesis}} and applied it to Stokes Q and U images, which had comparable noise levels ($<7.5~{\rm mJy~PSF^{-1}}$)\footnote{The noise threshold of $7.5~{\rm mJy~PSF^{-1}}$ is estimated based on the noise characteristics in Stokes QU datacubes of Cycle 2 observations. It corresponds to the mean value of it plus six times its variations measured by the standard deviation at frequencies not affected by the RFI.} in the frequency data cube of each observation. The frequency channels with noise $>7.5~{\rm mJy~PSF^{-1}}$ were flagged. The resulting Faraday cubes covered Faraday depths from $-50$ to $+50~{\rm rad~m^{-2}}$ in $0.25~{\rm rad~m^{-2}}$ steps, given the expected Faraday depth range of the observed emission in this field \citep[from $-10$ to $+13~{\rm rad~m^{-2}}$,][]{jelic14}. The resolution in Faraday depth was $\delta \Phi = 0.9~{\rm rad~m^{-2}}$ for all observations. The side lobes of the RMSF in Cycle 4 data were higher than in Cycle 2 data (see Fig.~\ref{fig:rmsf}) due to the gaps at frequencies contaminated by the broad RFIs (see Fig.~\ref{fig:noise_study}). Because the resolution in Faraday depth is comparable to the maximum detectable Faraday scale ($\Delta \Phi_{\rm scale} = 1.1~{\rm rad~m^{-2}}$), we are only sensitive to Faraday-thin structures ($\lambda^2\Delta\Phi_{\rm scale}\ll1$) or the edges of Faraday-thick structures \citep[$\lambda^2\Delta\Phi_{\rm scale}\gg1$][]{brentjens05}.

The noise in the Faraday cubes for the different observations is given in Table~\ref{table:misalignment}. The noise was estimated as the standard deviation of an image given in the polarised intensity at Faraday depth of $-50~{\rm rad~m^{-2}}$ and multiplied by a factor of $\sqrt{2}$. At this Faraday depth, we do not observe any polarised emission, and the image is dominated by noise. The factor $\sqrt{2}$ addresses the Rician distribution of the noise in the polarised intensity, which roughly corresponds to a normally distributed noise in Stokes Q and U \citep[e.g.][]{brentjens05}. A mean value of the noise in Cycle 2 observations is $91\pm10~{\rm \mu Jy~PSF^{-1}~RMSF^{-1}}$ and in Cycle 4 observations is $121\pm26~{\rm \mu Jy~PSF^{-1}~RMSF^{-1}}$. Higher noise in Faraday cubes of Cycle 4 data is due to a larger number of frequency channels in this cycle affected by RFI (see Fig.~\ref{fig:noise_study}). Observation 014 has the lowest noise among both Cycle 2 and 4 observations, and observation 021 has the lowest noise among Cycle 4 observations. Hence, we choose the 014 observation as a reference for Cycle 2. For Cycle 4 we take for consistency the same reference observation (024) as in \citet{ruiz21}, which is our second-best observation in terms of the noise in this cycle. 
We cannot choose the same reference observation for both cycles because of their different frequency configurations.

\begin{table} [t!]
\centering   
\begin{tabular}{ccccc}                
\hline \hline
ID & Cycle & Noise & $\Delta \Phi_\mathrm{shift}$\\
 & & [$\rm{\mu Jy\,PSF^{-1}\,RMSF^{-1}}$] & [$\rm{rad/m^{2}}$]\\ \hline 
009  & 2 & 106 & $-0.159 \pm 0.007$ \\  
010 & 2 & 94 & $-0.065 \pm 0.005 $\\  
011 & 2 & 88 & >1 \\  
012 & 2 &  84 & $0.053 \pm 0.007$\\  
013 & 2 & 88 & $0.002 \pm 0.006$\\  
014 & 2 &  82 & 'reference'\\  
015 & 2 & 84 & $0.010 \pm 0.005$ \\  
016 & 2 &  86 & $0.003 \pm 0.005$ \\  
017 & 2 & 85 & $0.010 \pm 0.006$\\ 
018 & 2 & 112 & $0.052 \pm 0.005$ \\ 
\hline
020 & 4 & 107  & $-0.033 \pm 0.005 $\\ 
021 & 4 & 91 & $-0.021 \pm 0.004$ \\ 
022 & 4 & 133 & $0.020 \pm 0.004$\\ 
023 & 4 & 111 & $0.011 \pm 0.004$\\ 
024 & 4 & 102 & 'reference' \\ 
026 & 4 & 112 & $0.033 \pm 0.005$\\ 
027 & 4 & 112 & $0.017 \pm 0.004$\\ 
028 & 4 & 179 & $0.045 \pm 0.006$ \\ 
030 & 4 & 164 & $0.114\pm 0.005$ \\ 
031 & 4 & 103 & $0.080\pm 0.004$ \\ 
032 & 4 & 111 & $0.101 \pm 0.004$\\ 
\hline
\end{tabular}
\caption{Calculated noise in the Faraday cubes given for different observations and their relative shift in Faraday depth ($\Delta\Phi_{\rm shift}$) with respect to the reference observation (calculated in Sect.~\ref{sub:ionoEFF}). An ID of each observation corresponds to the one given in Table 1 in \citet{sabater21}.} 
\label{table:misalignment}
\end{table}

We used publicly available code \texttt{rmclean3d} from \texttt{RM-Tools}\footnote{\url{https://github.com/CIRADA-Tools/RM-Tools}} \citep{RMTools} to deconvolve the Faraday cubes for the side lobes of the RMSF. The code is based on RM-CLEAN algorithm described in \citet{heald09}. We used a threshold of five times the noise in the Faraday cube during the RM-CLEAN process.

\subsection{Comparison with a previous LOFAR commissioning observation} \label{sub:Galsingle}
The ELAIS-N1 field was observed previously with LOFAR during its commissioning phase \citep{jelic14}. That observation was done in a limited frequency range from 138 MHz to 185 MHz. Here we make a comparison between that observation and observations used in this work. The comparison is done using Faraday cubes in the polarised intensity. 

Noise in a Faraday cube of the commissioning observation (a single 8h synthesis) was $300~{\rm \mu Jy~PSF^{-1}~RMSF^{-1}}$ \citep{jelic14}. This is around $3.6$ times higher than the noise in the individual Faraday cubes presented in this work. The difference arises from the limited available frequency bandwidth during the commissioning phase of LOFAR and the use of a simpler calibration strategy that addressed only direction-independent effects.

The commissioning observation of the ELAIS-N1 field revealed polarised diffuse emission over a wide range of Faraday depths ranging from $-10$ to $+13~{\rm rad~m^{-2}}$ \citep{jelic14} given a resolution of $1.75~{\rm rad~m^{-2}}$ in Faraday depth. The most prominent features of that emission are seen in the left image of Fig. 7 in \citet{jelic14}, showing the highest peak value of the Faraday depth spectrum at each pixel (RA, Dec). The mean surface brightness of that emission is $2.6~{\rm mJy~PSF^{-1}~RMSF^{-1}}$. The same figure also shows the Faraday depth of each peak in an image presented on the right. 

\begin{figure*}[t!]
   \centering
   \includegraphics[width=0.45\linewidth]{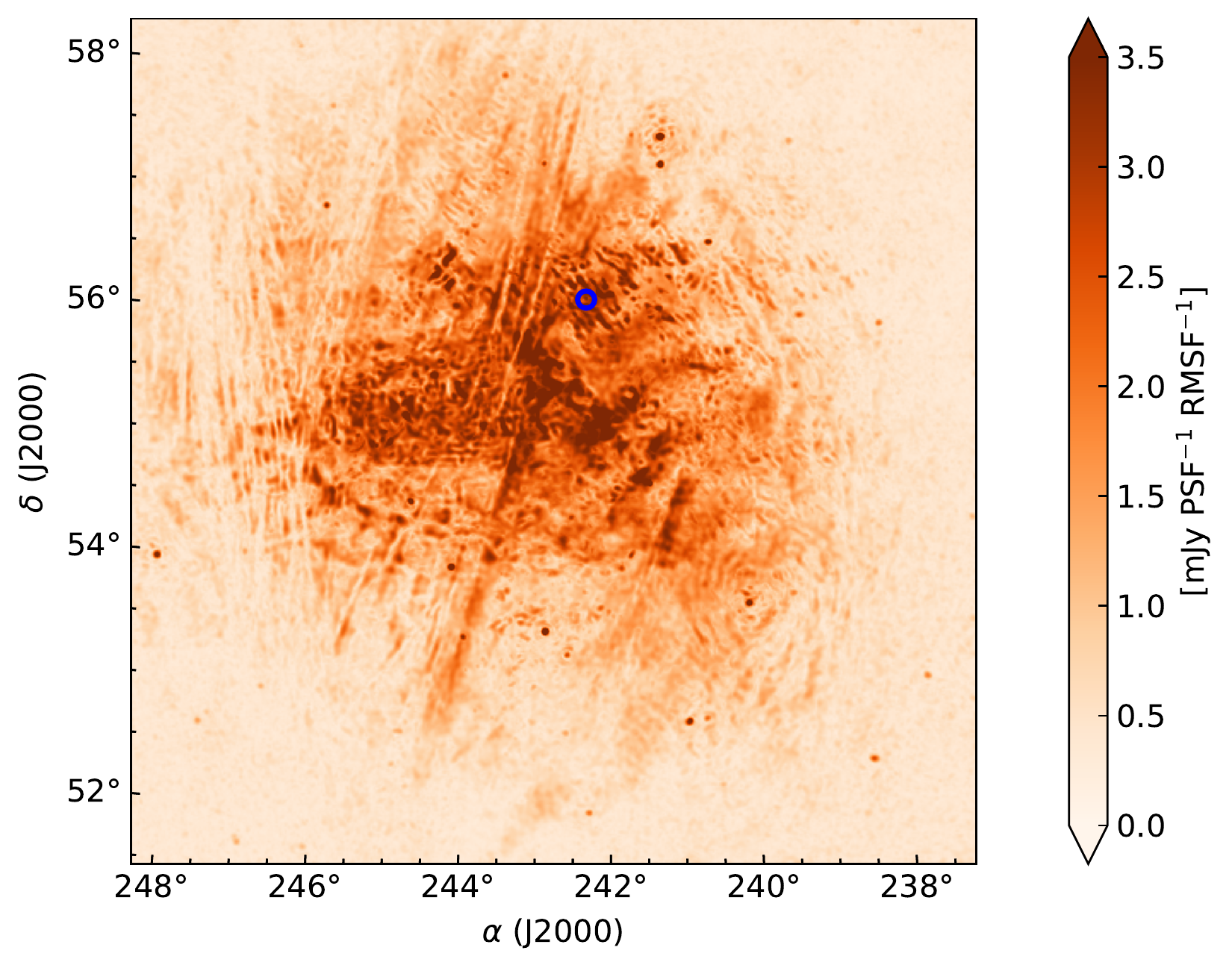} \includegraphics[width=0.45\linewidth]{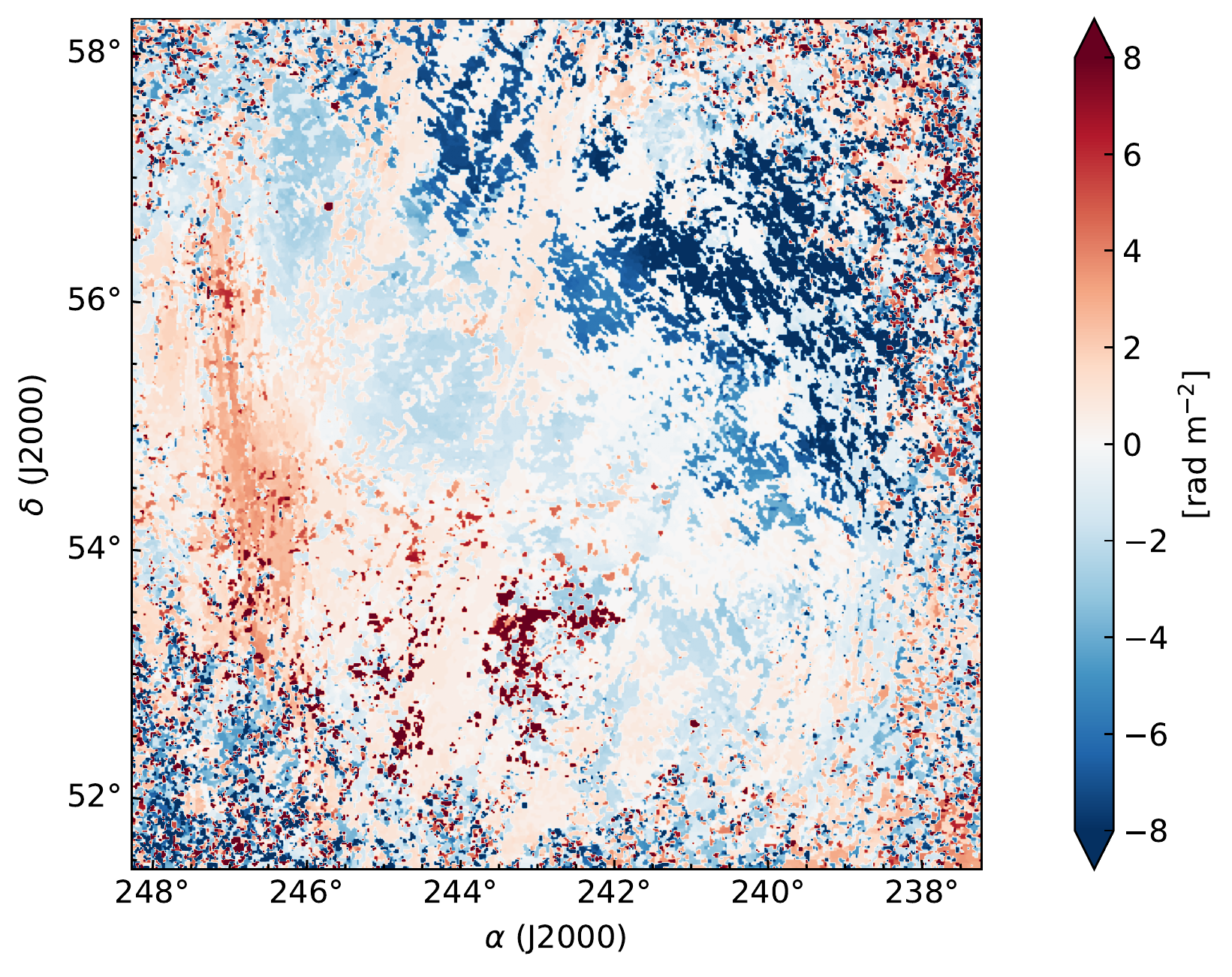}
      \caption{Image of the highest peak of the Faraday depth spectrum in the polarised intensity (left) and a corresponding image of a Faraday depth of the highest peak (right) for the observation with the lowest noise (014, the reference observation for Cycle 2). The blue circle in the left image marks a randomly chosen location for which a Faraday spectrum is presented in Fig.~\ref{fig:misalignment009_014}.}
         \label{fig:Max_int_014}
\end{figure*}

We constructed the same images for the observations analysed in this work. The images for the observation that has the lowest noise level (014) are presented in Fig.~\ref{fig:Max_int_014} as an example. Images for all other observations are very similar to these. The observed diffuse emission in the left image of Fig.~\ref{fig:Max_int_014} shows morphological similarity with the one detected in the commissioning observation \citep[see left image in Fig. 7 in][]{jelic14}. The observed morphological features appear much sharper despite comparable angular resolution in both observations. This is due to almost two times better resolution in Faraday depth than in the commissioning observation. As a consequence, the observed emission suffers less from depolarisation, as is the case,  for example, for a filamentary structure oriented north-south in the central part of the image. The filament is depolarised in the commissioning observation, while it is visible in observations presented in work. Due to a better signal-to-noise ratio, there is also more emission visible towards the edges of the image, where the emission is attenuated by the LOFAR primary beam. The mean surface brightness of the observed emission in the central part of the image is $3.0~{\rm mJy~PSF^{-1}~RMSF^{-1}}$, which is a bit brighter than in the commissioning observation. The emission appears in a range of Faraday depths from $-16$ to $+14~{\rm rad~m^{-2}}$, starting at slightly smaller and ending at slightly larger Faraday depths than in the commissioning observation. Further discussion on characteristics of the observed emission are in Sect.~\ref{sub:diffuse}.

\section{Analysis of a relative shift in Faraday depth between different observations} \label{sub:ionoEFF}

Data analysed in this work were corrected for the ionospheric Faraday rotation by the satellite-based TEC measurements (see Sect.~\ref{sub:deep_filed_observations}). To check how well the data are corrected, we make a relative comparison between each observation and the reference observation by cross-correlating Faraday cubes. 

Instead of explicitly cross-correlating Faraday cubes of two observations ($a$ and $b$) as in \citet{jelic15}, it is computationally more efficient to use the Fourier transform’s cross-correlation property. We calculated the cross-correlation function by effectively performing RM-synthesis on $P_a(\lambda^{2})P_b^\ast(\lambda^{2})$, as proposed by \citet{brentjens18} and implemented in the above-mentioned publicly available code \texttt{rm-synthesis}.  Using this code, we evaluate the cross-correlation function for Faraday depths between $-5$ and $+5~{\rm rad~m^{-2}}$ in  $0.01~{\rm rad~m^{-2}}$ steps. We expect the relative shift between the observations to be $\lesssim 1~{\rm rad~m^{-2}}$. 

\begin{figure*}[t!]
  \centering
  \includegraphics[width=0.49\linewidth]{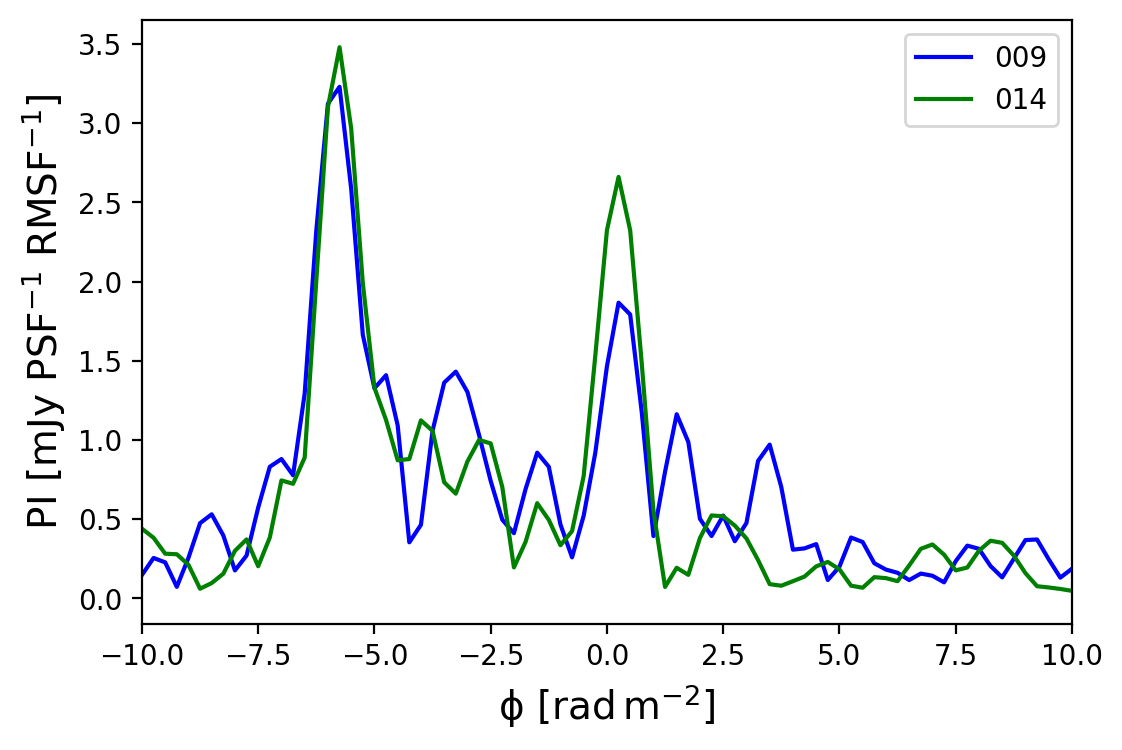}
  \includegraphics[width=0.49\linewidth]{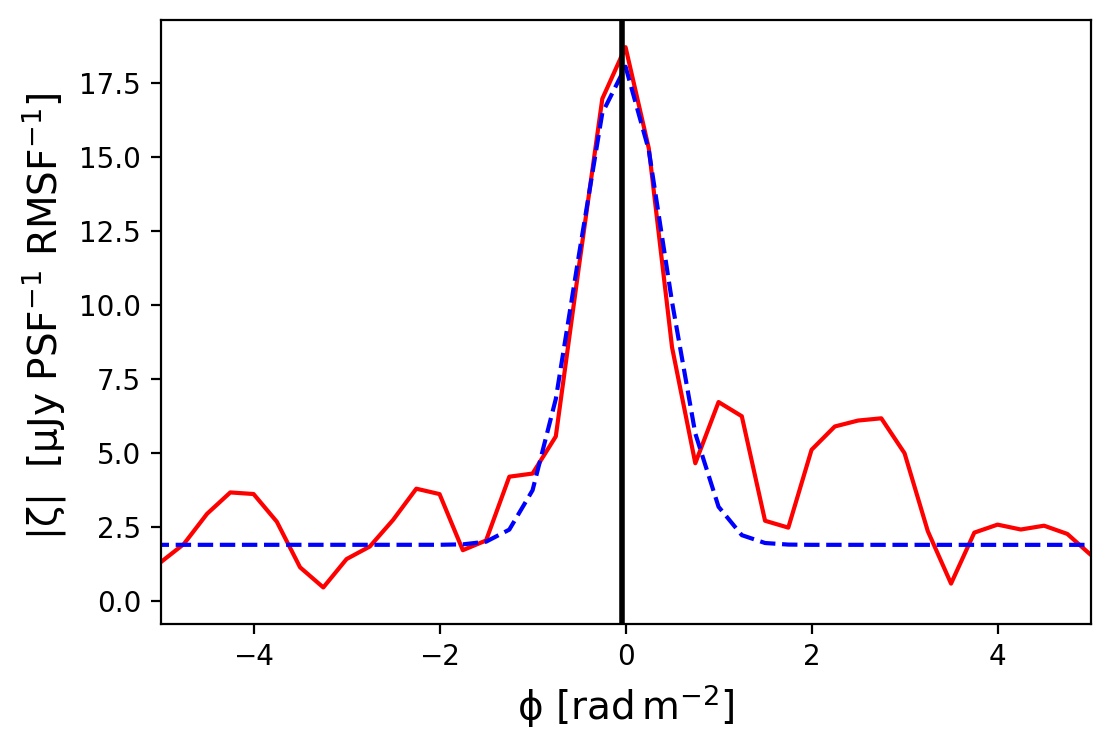}
  \caption{Example of a Faraday spectrum given in the polarised intensity for the 009 observation (blue line) and the reference (014) observation at a randomly chosen location marked with a blue circle in Fig.~\ref{fig:Max_int_014} (RA $242^\circ18'03.60"$ and Dec $56^\circ08'16.80"$), given on the left panel. Calculated modulus of the complex cross-correlation function $|\mathrm{\zeta}|$ (red line) for the given Faraday spectra, fitted with a Gaussian (blue dashed line) to estimate the misalignment between the two observations (black vertical line) at this specific location, given on the right panel.}
   \label{fig:misalignment009_014}
\end{figure*}

We illustrate the applied method in Fig.~\ref{fig:misalignment009_014} by giving examples of Faraday spectra in the polarised intensity for two observations (009 and 014; left image) and the modulus of their evaluated complex cross-correlation function $\zeta$ (right image). Faraday spectra are taken for a random (RA, Dec) pixel in the cube (marked with a blue circle in Fig.~\ref{fig:Max_int_014}), where the observed emission is relatively bright. To find a Faraday depth of the cross-correlation function's peak, we fitted a Gaussian to the peak. At this specific position in the image, the Faraday spectrum of the 009 observation is shifted by $-0.047\pm0.006~{\rm rad~m^{-2}}$ with respect to a reference (014) observation. 

To find a common shift in Faraday depth across the FoV, we averaged the complex cross-correlation functions for all pixels (RA, Dec), where the observed emission in a reference observation has the highest peak value of the Faraday depth spectrum at least ten times larger than the noise ($\geq 82~{\rm mJy~PSF^{-1}~RMSF^{-1}}$). This will improve the signal-to-noise ratio and therefore the location of the main peak of the cross-correlation function. When averaging, we assume that the variation of shifts in Faraday depth across the FoV is much smaller than the width of the main peak of the RMSF ($\ll0.9~{\rm rad~m^{-2}}$).

\begin{figure}[h!]
  \centering
  \includegraphics[width=0.5\textwidth]{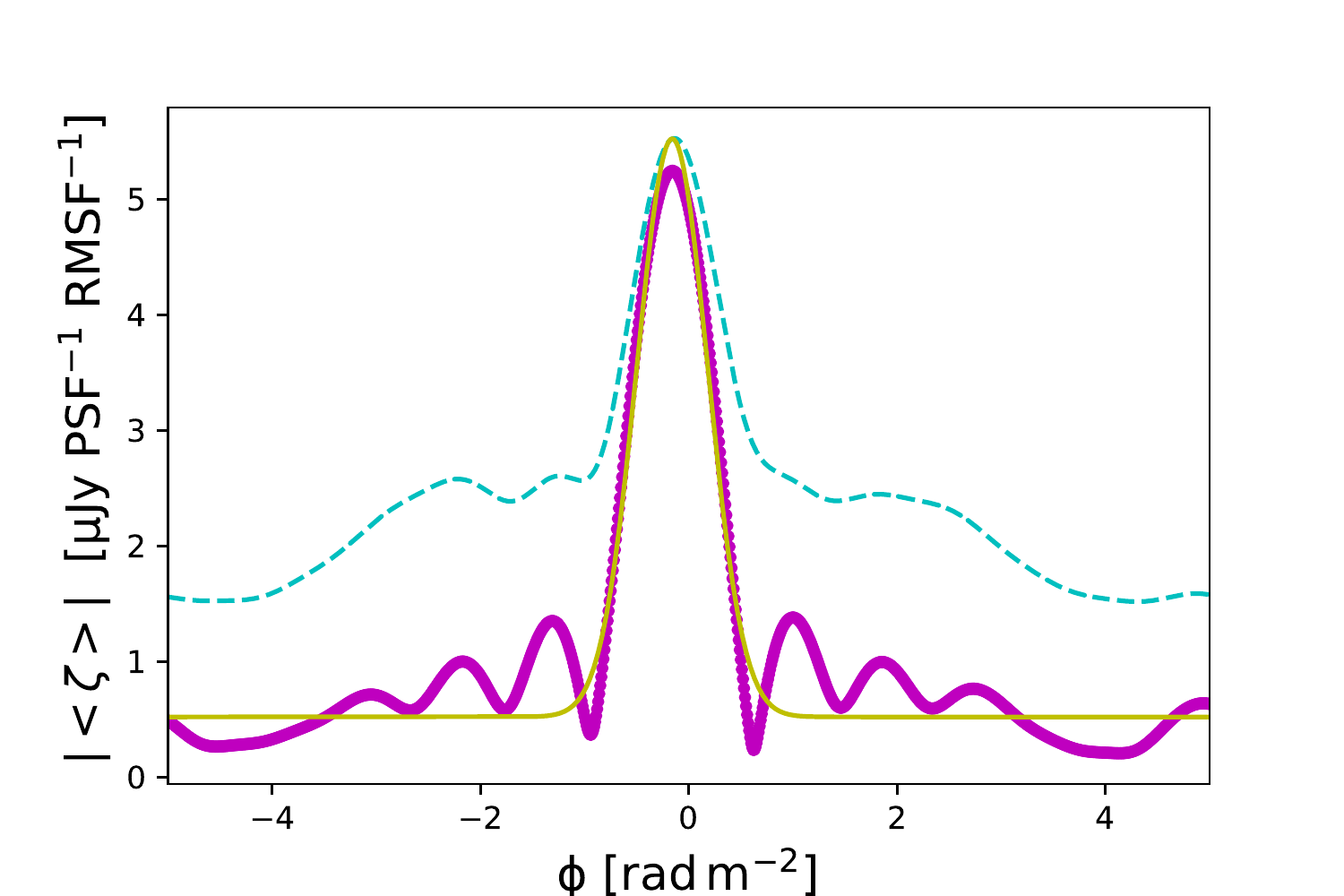}
  \caption{Calculated modulus of the averaged complex cross-correlation function for observations 009 and 014
(solid magenta line) and variations of the cross-correlation function across the Fov as measured
by a standard deviation (dashed cyan line). A misalignment between two observations is determined by fitting a Gaussian to the peak (solid yellow line).}
   \label{fig:gauss_fit009_014}
\end{figure}

Figure~\ref{fig:gauss_fit009_014} shows the calculated modulus of the averaged complex cross-correlation function for observations 009 and 014 (magenta solid line). The same figure also gives variations of the cross-correlation function across the FoV as measured by a standard deviation (cyan dashed line). We then fit a Gaussian to the peak and find that the 009 observation is shifted by $-0.159\pm0.007~{\rm rad~m^{-2}}$ with respect to the reference observation. 

Calculated relative shifts in Faraday depth for all other observations are given in Table~\ref{table:misalignment}. There is no significant difference in misalignment between the two cycles. The observations are on average misaligned by $\pm(0.046\pm 0.042)~{\rm rad~m^{-2}}$ with respect to the reference observation. The only exception is 011 observation, which shows a misalignment larger than $1~{\rm rad~m^{-2}}$ and is analysed in detail in Appendix~\ref{app:restoring}.

The estimated misalignments are comparable to the one found in the analysis of five LOFAR observations of the 3C~196 field \citep[$0.1\pm0.08~{\rm rad~m^{-2}}$;][]{jelic15}. This verifies the reliability of ionospheric Faraday rotation corrections estimated using the satellite-based TEC measurements. The related uncertainties are mostly connected to daily systematic biases in the TEC measurements of $\sim1~{\rm TEC}$ unit, translating to an error in the ionospheric rotation measure of $\sim0.1~{\rm rad~m^{-2}}$. The misalignment for observations 020, 027, 028, 030, and 031 are in agreement with the one estimated by \citet{ruiz21} within the errors. \citet{ruiz21} based their analysis using a single Faraday spectrum at the location of the peak pixel of the reference polarised source, while we used all pixels that show bright polarised diffuse emission. Therefore, estimated errors are $\sim5$ times smaller in our work than in their work.

\section{Stacking very low-resolution data}
\label{sub:stack}
To stack images of different observations together, we first need to `de-rotate' the observed polarisation angle of each observation by its estimated shift with respect to the reference observation ($\Delta\Phi_\mathrm{shift}$, see Table~\ref{table:misalignment}). We multiplied the complex polarisation $P_i(\lambda^2)=Q_i(\lambda^2) + \mathrm{i}U_i(\lambda^2)$ given at each wavelength (frequency) by $\exp^{\mathrm{i}2\Delta\Phi_\mathrm{shift}\lambda^{2}}$:
\begin{equation}
\tilde{P}_i(\lambda^2)=\tilde{Q}_i(\lambda^2)+\mathrm{i}\tilde{U}_i(\lambda^2)=\left(Q_i(\lambda^2)+\mathrm{i}U_i(\lambda^2)\right)\exp^{\mathrm{i}2\Delta\Phi_\mathrm{shift}\lambda^{2}}.
\end{equation}
This way, the correction is applied to the whole Faraday spectrum simultaneously.

\begin{figure}
    \centering
    \includegraphics[width=\linewidth]{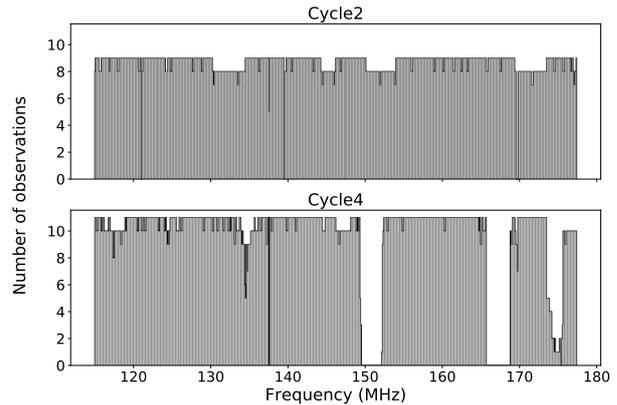}
    \caption{Number of observations per frequency channel used in the stacked data cube from Cycle 2 (upper plot) and Cycle 4 (lower plot) data.  Although the data cover the same frequency range, they have different frequency configurations and hence a different number of frequency channels (see Sect.~\ref{sub:deep_filed_observations}).}
    \label{fig:freqCov}
\end{figure}

We then stack all corrected images of each observing cycle by calculating the weighted average at each wavelength (frequency): 
\begin{equation}\label{eq:weight}
   P_\mathrm{combined}(\lambda^2)=\frac{\sum_i \tilde{Q}_i(\lambda^2)w^{\tilde{Q}}_i(\lambda^2)}{\sum_i w^{\tilde{Q}}_i(\lambda^2)}+\mathrm{i}\frac{\sum_i \tilde{U}_i(\lambda^2)w^{\tilde{U}}_i(\lambda^2)}{\sum_i w^{\tilde{U}}_i(\lambda^2)},
\end{equation}
where $w^{\tilde{Q},\tilde{U}}_{i}(\lambda^2)$ is a wavelength- (frequency) dependent weight for each observation defined as the inverse of the variance of the noise in Stokes Q and U images. We recall that the noise in Stokes Q and U were comparable and were calculated in the corner of each image where the polarised emission was not present. We are not able to stack the data from two cycles directly because of their different frequency channel widths (see Sect.~\ref{sub:deep_filed_observations}). They are combined at a later stage in Faraday depth. Figure~\ref{fig:freqCov} shows a number of images per frequency channel used in the final stacked data cube for Cycle 2 and Cycle 4. There are on average nine images added per frequency channel in Cycle 2 and 11 images in Cycle 4. 

Images of observation 011 are not used for the stacked data cube of Cycle 2 because of their relatively poor quality compared to the images of all other observations. This choice does not have any significant impact on the final result. Once the data of each observing cycle were stacked, we applied the RM synthesis.

The noise in the stacked Faraday cube of Cycle 2 data is 32~${\rm \mu Jy~PSF^{-1}~RMSF^{-1}}$ and of Cycle 4 data is 40~${\rm \mu Jy~PSF^{-1}~RMSF^{-1}}$. In both cases, this is $\sim 3$ times less than the mean value of noise in Faraday cubes of individual observations ($\rm{91\pm10}$~${\rm \mu Jy~PSF^{-1}~RMSF^{-1}}$ and $\rm{121\pm26}$~${\rm \mu Jy~PSF^{-1}~RMSF^{-1}}$, respectively). The noise in the stacked Faraday cube is reduced by the square root of the number of observations that are stacked, as expected.

We calculated the cross-correlation between the stacked Faraday cubes of two cycles as a function of the displacement in Faraday depth to check for their alignment. We only considered Faraday spectra that have a peak flux in the polarised intensity $\geq 82~{\rm mJy~PSF^{-1}~RMSF^{-1}}$, the same limit as the one used in Sect.~\ref{sub:ionoEFF}. The resulting cross-correlation functions were then averaged, and their common peak was fitted with a Gaussian. The two cubes are aligned in Faraday depth within the error of the fit and can be combined directly to the final Faraday cube. 

We combined the stacked Faraday cubes of two cycles by calculating the weighted average in Faraday depth:
\begin{equation}\label{eq:weight}
   P_\mathrm{combined}(\Phi)=\frac{\sum_i Q_i(\Phi)w^{\Phi}_i}{\sum_i w^{\Phi}_i}+\mathrm{i}\frac{\sum_i U_i(\Phi)w^{\Phi}_{i}}{\sum_i w^{\Phi}_i},
\end{equation}
where $w^{\Phi}_i$ is a Faraday depth independent weight for each Faraday cube defined as the inverse of the variance of the noise in Stokes Q and U Faraday cubes. The noise in the Faraday cube is estimated as the standard deviation of an image given in Stokes Q and U at $-50~{\rm rad~m^{-2}}$. This is the Faraday depth, where we do not observe any polarised emission and the image is dominated by noise.

\begin{figure*}[h!]
  \centering
  \includegraphics[width=0.3\linewidth]{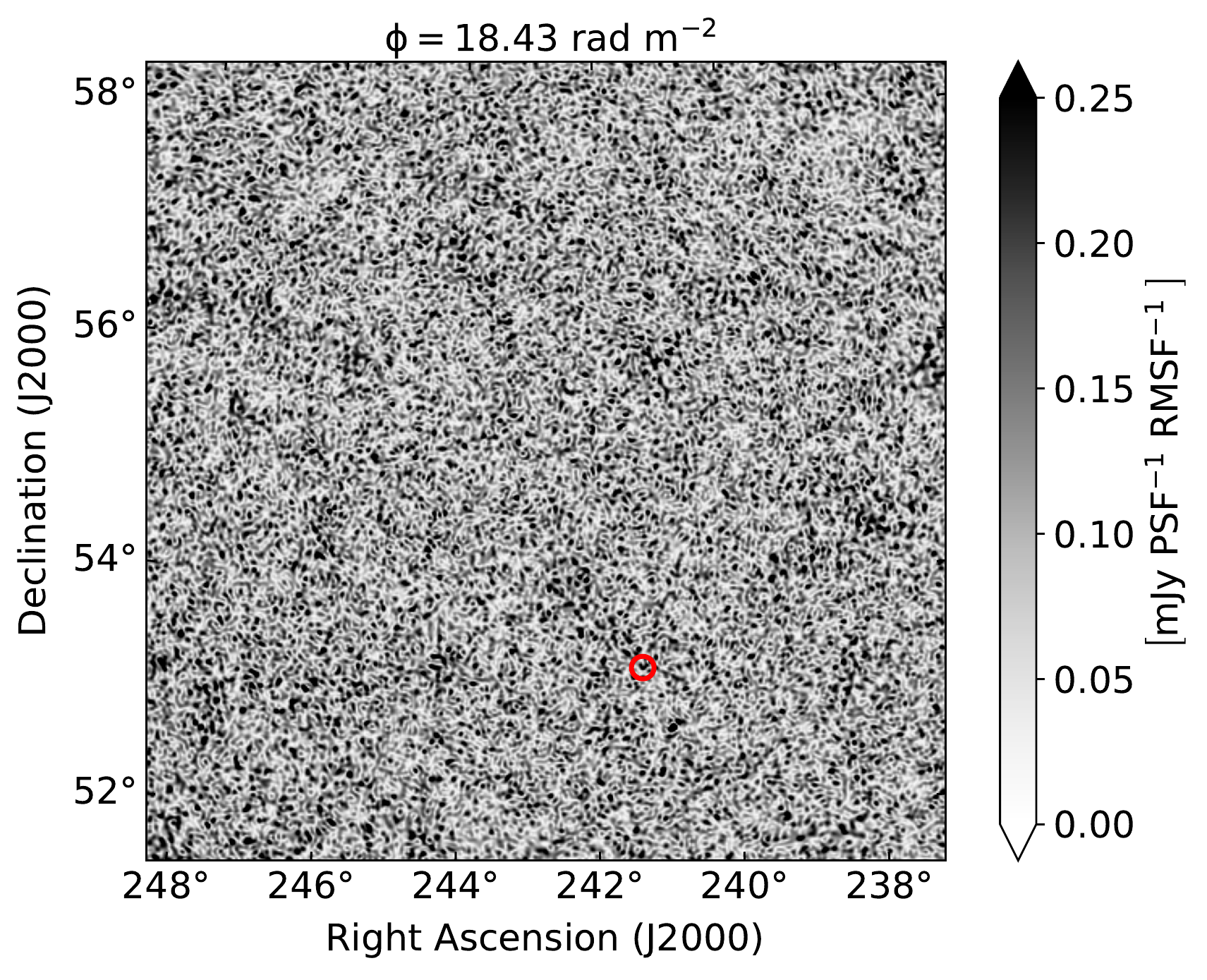}
  \includegraphics[width=0.3\linewidth]{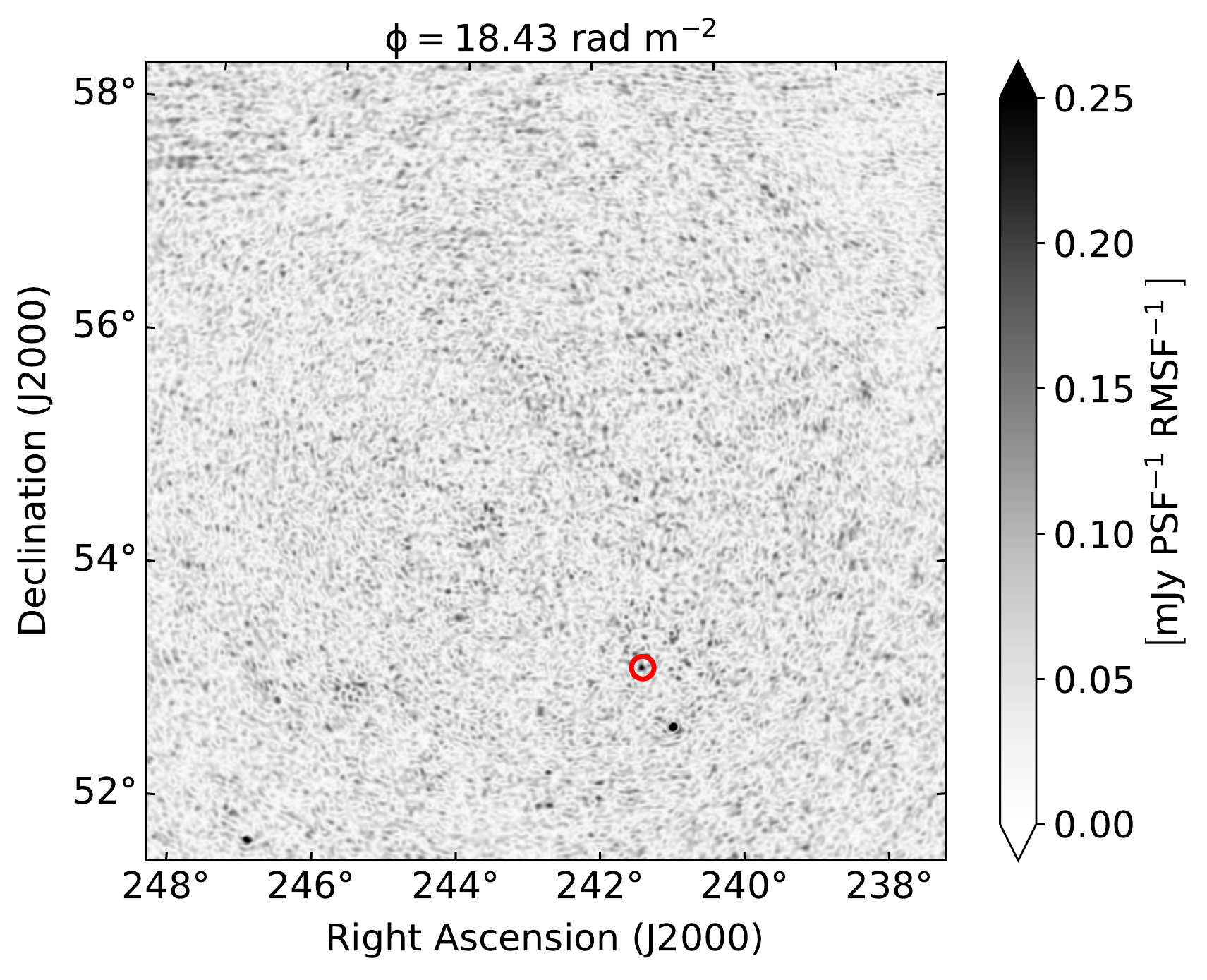}
  \includegraphics[width=0.33\linewidth]{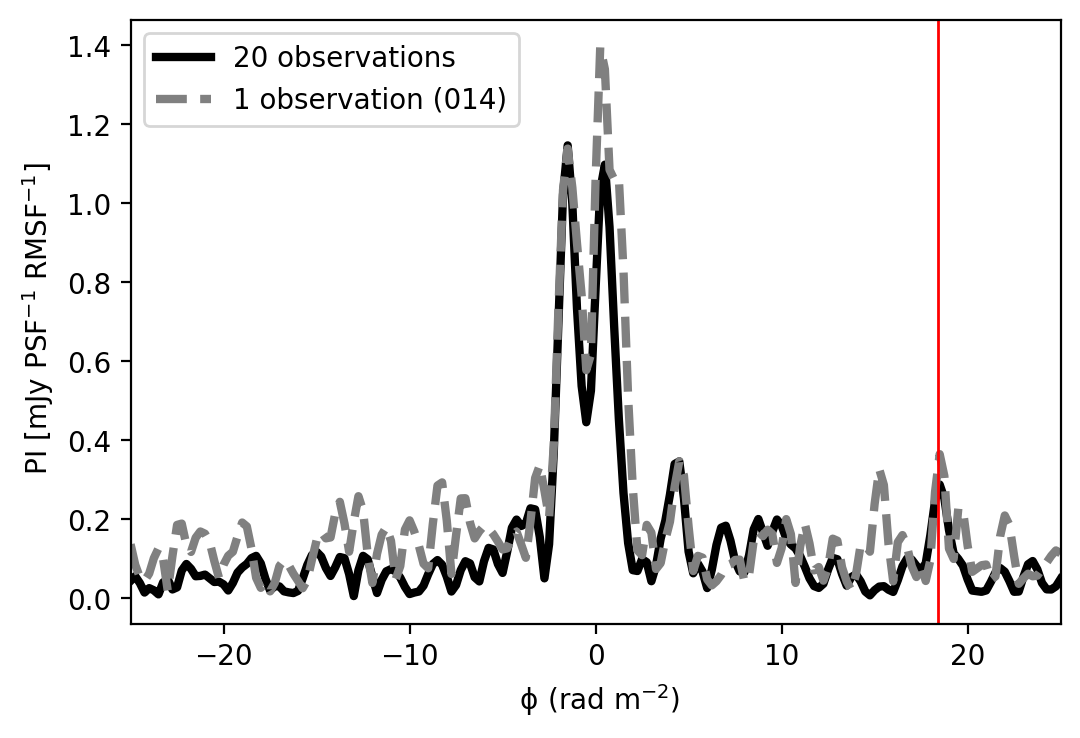}
  \includegraphics[width=0.3\linewidth]{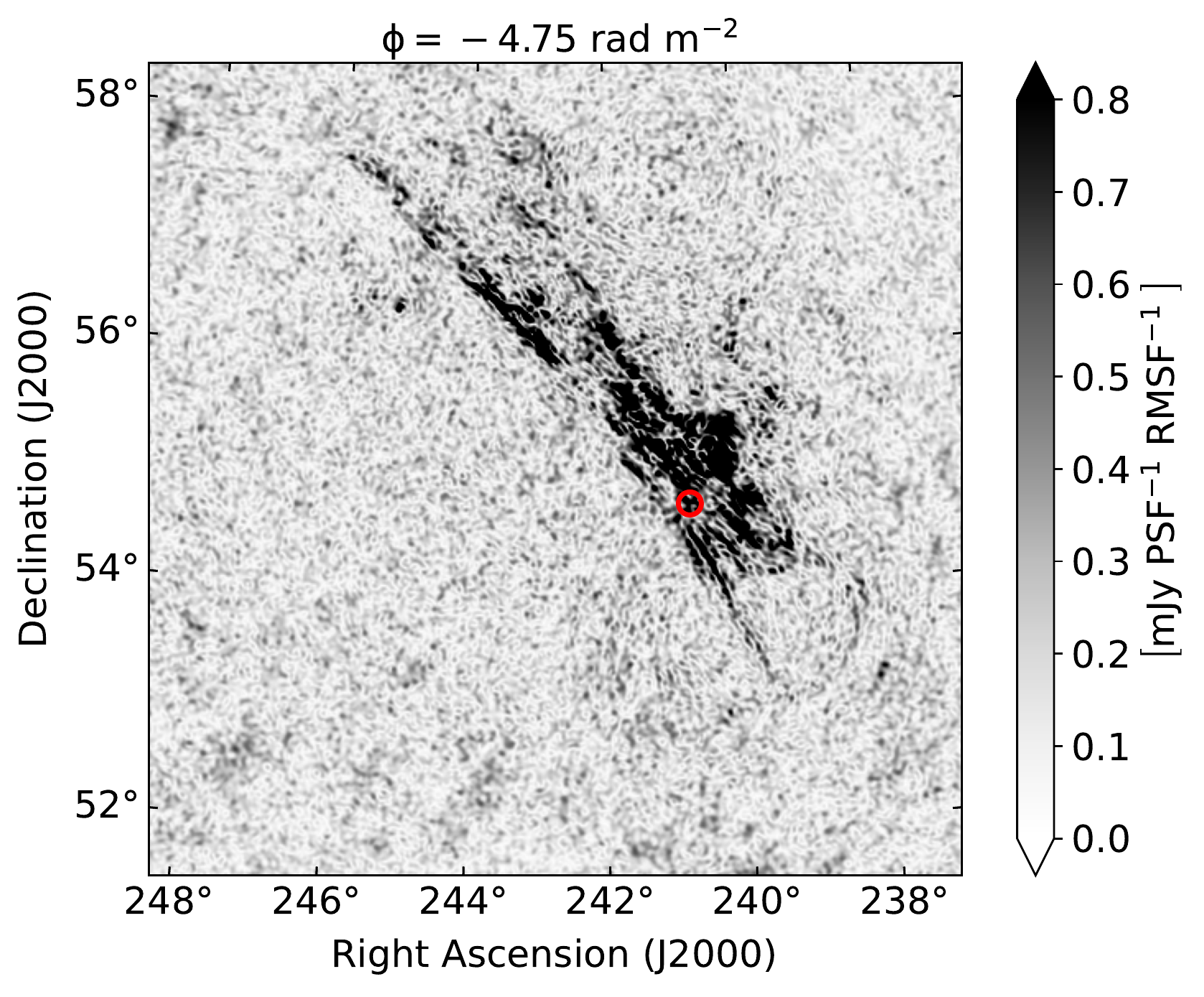}
  \includegraphics[width=0.3\linewidth]{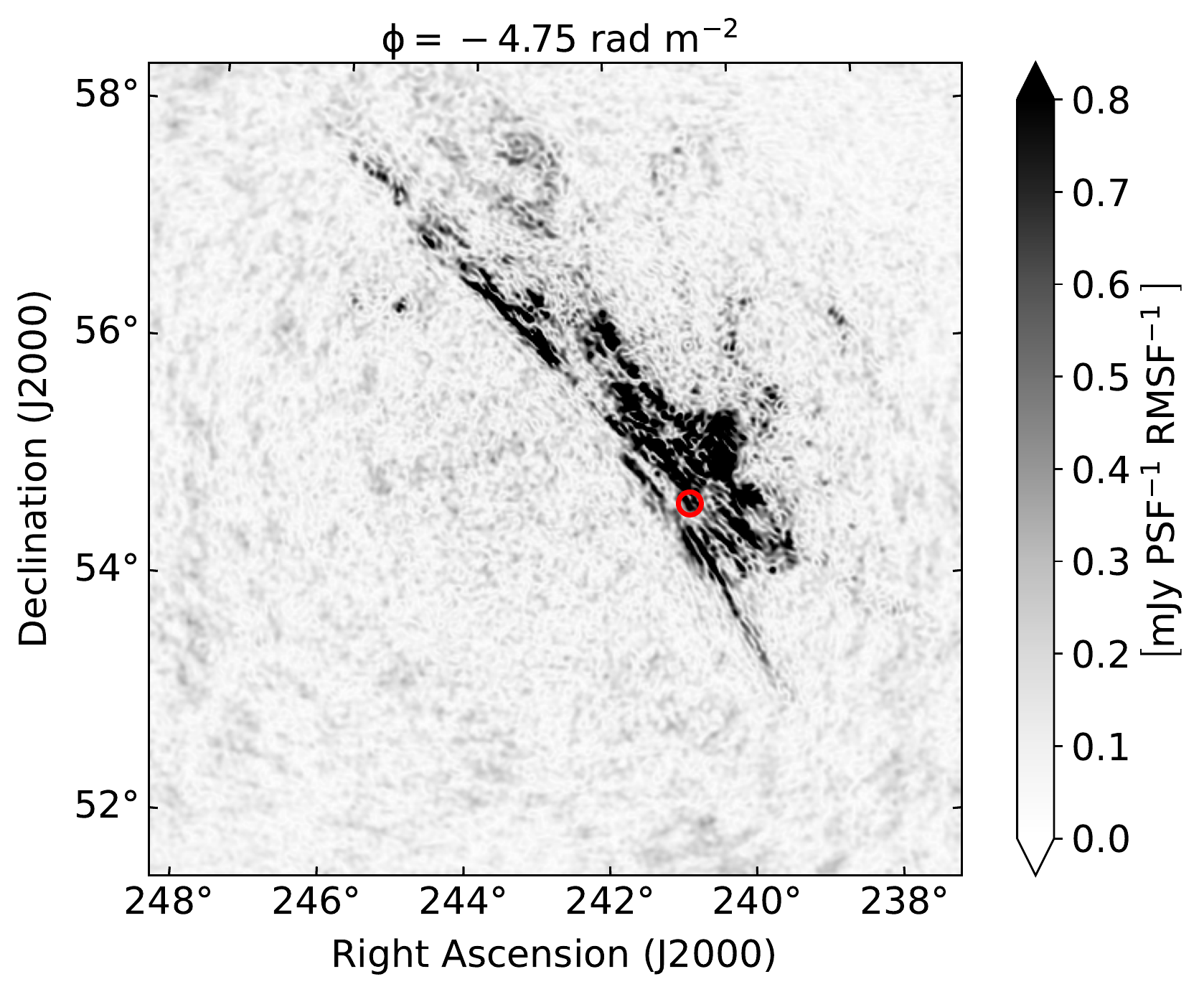}
  \includegraphics[width=0.33\linewidth]{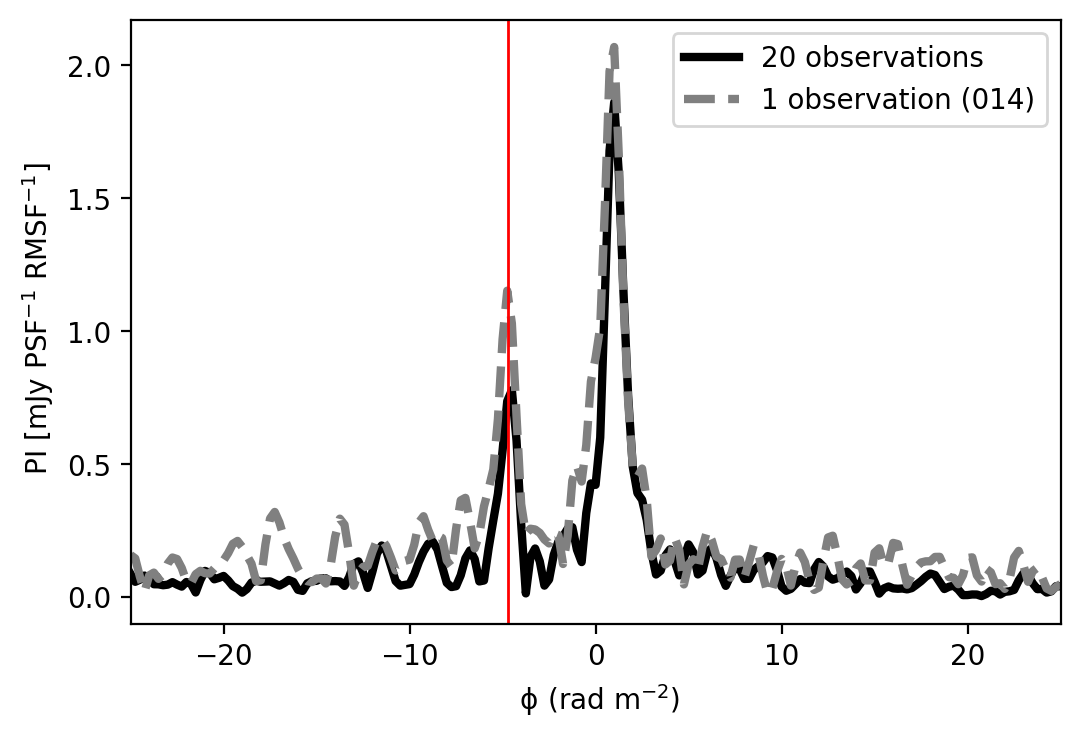}
  \caption{Example of a successful (upper panels) and an unsuccessful detection of a polarised source (lower panels) in the presented Faraday cubes \citep[sources with ID 10 and 07 in][respectively]{ruiz21}. Polarised intensity images in the reference (014, left images) and in the final stacked Faraday cube (middle images) are given at the closest available Faraday depth, such as that of the source. The location of the source in each image is marked with the red circle, while the corresponding Faraday spectra are given in plots on the right. A reported Faraday depth of the sources by \citet{ruiz21} are marked with vertical red lines.}
\label{fig:polsource_examples}
\end{figure*}

\section{The final stacked Faraday cube}
The final stacked Faraday cube combines images of 20 ELAIS-N1 LoTSS-Deep Fields observations, $\sim150$ hours of data in total. The cube covers Faraday depths from $-50$ to $+50~{\rm rad~m^{-2}}$ in $0.25~{\rm rad~m^{-2}}$ steps. The resolution in Faraday depth is $0.9~{\rm rad~m^{-2}}$, as defined by the resolution of the stacked Faraday cubes of the two observing cycles. The final image noise is 27~${\rm \mu Jy~PSF^{-1}~RMSF^{-1}}$, which is $\sim\sqrt{20}$ smaller than the mean value of noise in Faraday cubes of every individual observation in the two cycles, as expected. In the following two subsections, we first cross-check if we detect the radio sources presented in the catalogue of \citet{ruiz21}, and then we analyse and discuss the observed diffuse Galactic polarised emission.

\subsection{Cross-checking the detection of the radio sources} 
\label{sub:radio_sources}
We used the catalogue of the polarised sources provided by \citet{ruiz21} to check how many of them we detect in our final stacked Faraday cube. The purpose of this comparison is only to verify our stacking method on very low-resolution data. It is not meant to provide an in-depth analysis of the polarised sources. This is done in \citet{ruiz21} and in a follow-up work using the high-resolution data ($20''$ and $6''$, respectively), which are better suited for such analysis than the very low-resolution data ($4.3'$) used in this work. 

We extracted the Faraday spectra and inspect the images in our final polarised intensity cube at locations of polarised sources provided in the catalogues. We have a clear detection of nine out of ten radio sources from \citet[][Table 2, ID 01--06 and 08--10]{ruiz21}, while one of them (ID 07) is difficult to identify due to the presence of the diffuse polarised emission in our Faraday cube. Two examples are given in Fig.~\ref{fig:polsource_examples} for sources with IDs 10 and 07. In the first example, the source is not contaminated by diffuse emission. There is a clear signature of it in the Faraday spectrum of the stacked data. This source is, however, difficult to detect in the reference observation due to a poorer signal-to-noise ratio than in the stacked data. In the second example, we don't find the signature of the source, either in the stacked data or in the reference observation, due to contamination by diffuse polarised emission that dominates the image and the Faraday spectrum at the location of the source. 

The rotation measures of successfully detected sources in our final cube are in agreement with the values provided in the catalogue, taking into account a resolution in Faraday depth of $0.9~{\rm rad~m^{-2}}$ and a difference in angular resolution of the used data. The polarised radio source catalogue is based on high-resolution LoTSS data ($20''$), while in our work we used very low-resolution LoTSS data ($4.3'$). Therefore, morphologies of polarised sources are mostly not resolved in our data. If a source was unresolved in our data, while in reality it has, for example, two lobes \citep[see a source with ID 07, Fig. 7 in][]{ruiz21} whose RMs do not differ more than a resolution of the data in Faraday depth, we observed its rotation measure as an averaged value of the two lobes and additionally weighted by their relative brightness.

\begin{figure*}[h!]
  \centering
  \includegraphics[width=0.33\linewidth]{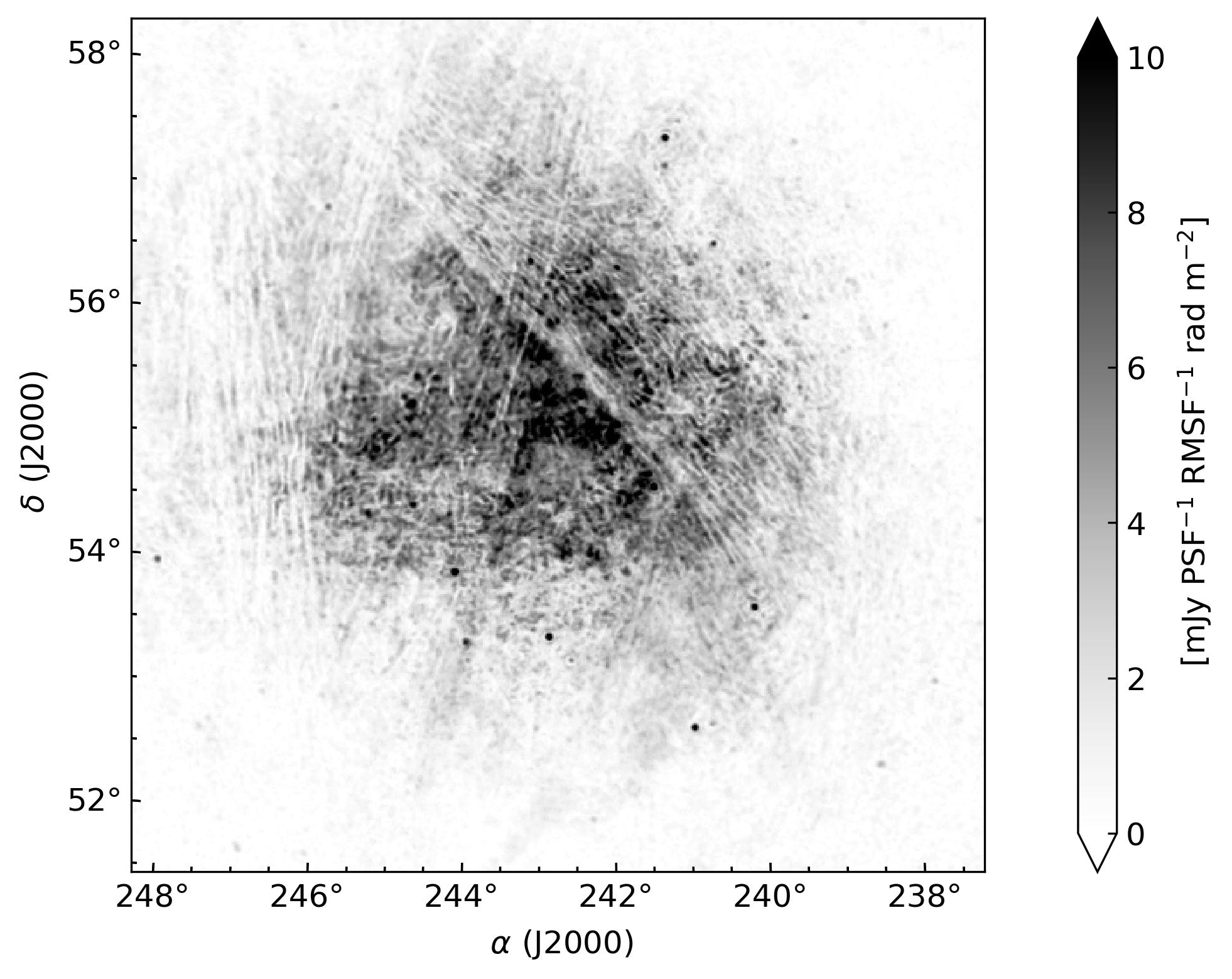}
  \includegraphics[width=0.33\linewidth]{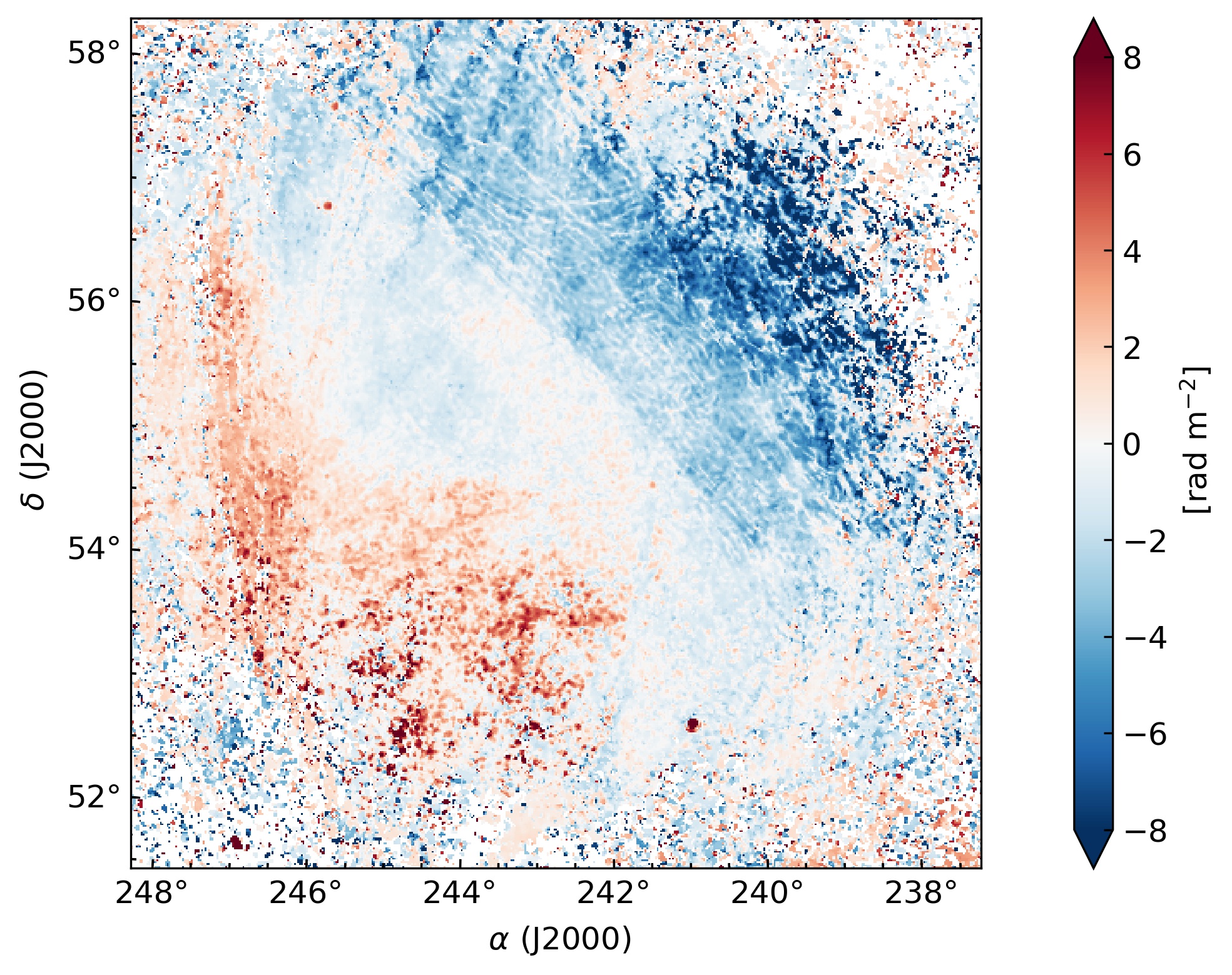}
  \includegraphics[width=0.33\linewidth]{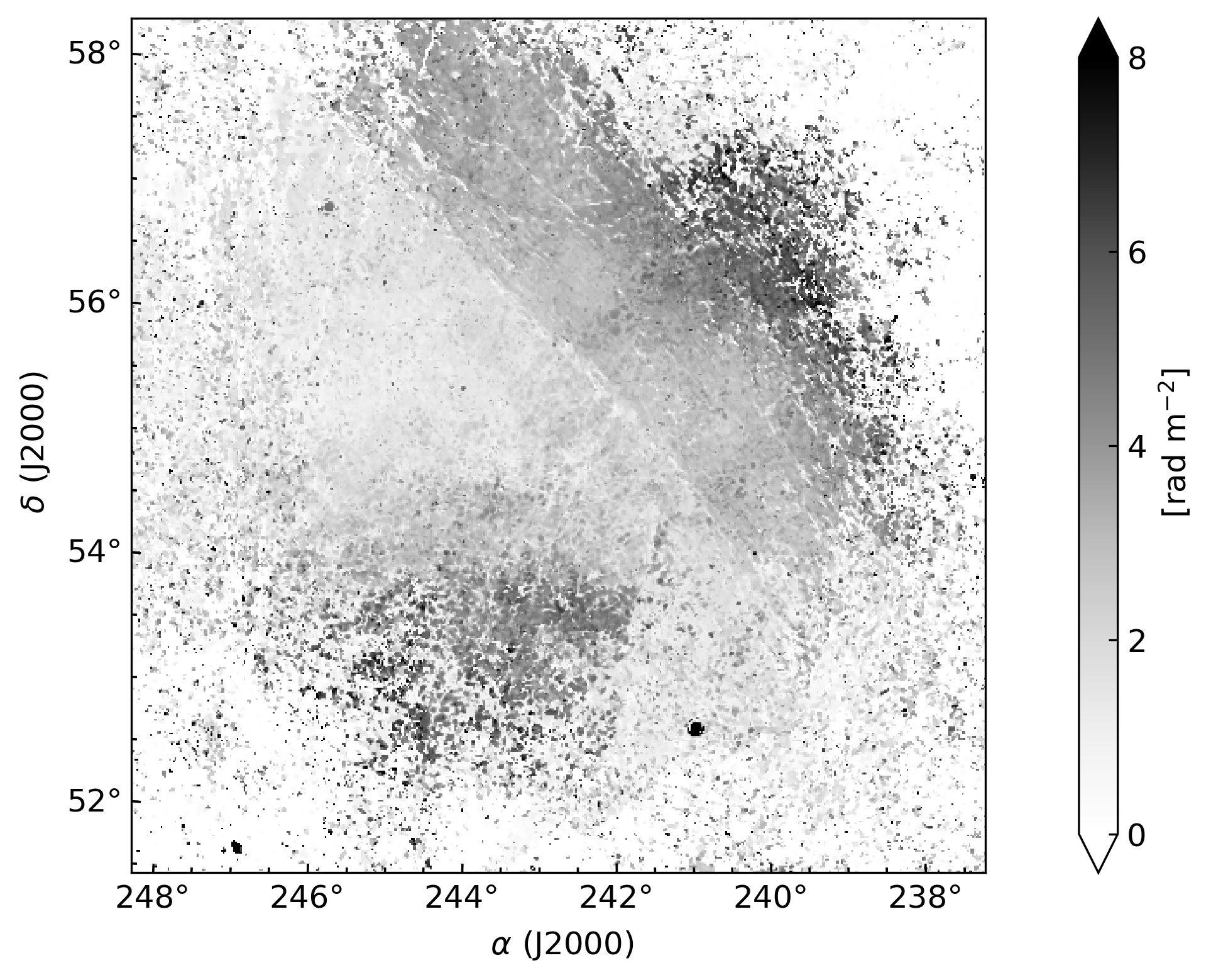}
  \includegraphics[width=0.33\linewidth]{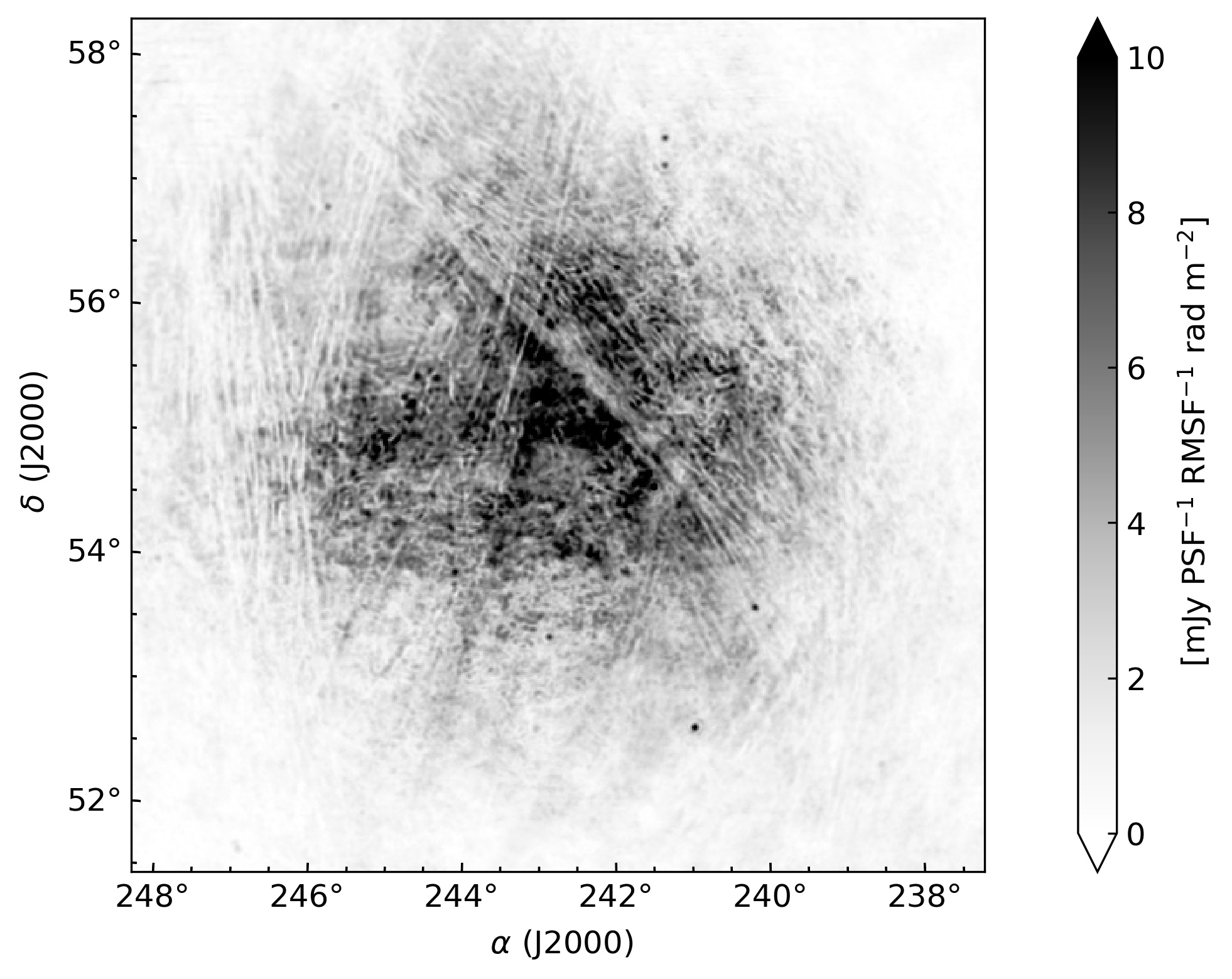}
  \includegraphics[width=0.33\linewidth]{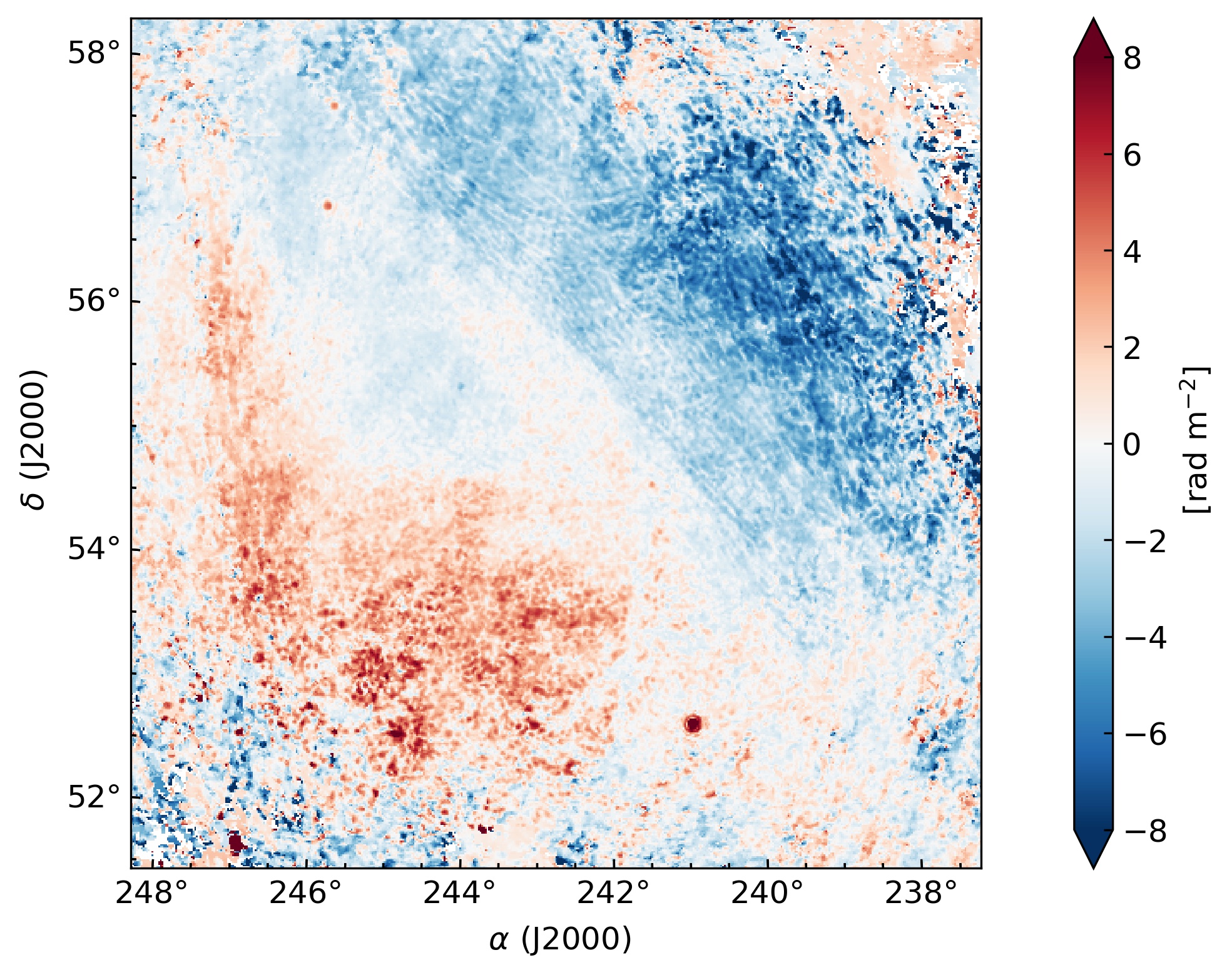}
  \includegraphics[width=0.33\linewidth]{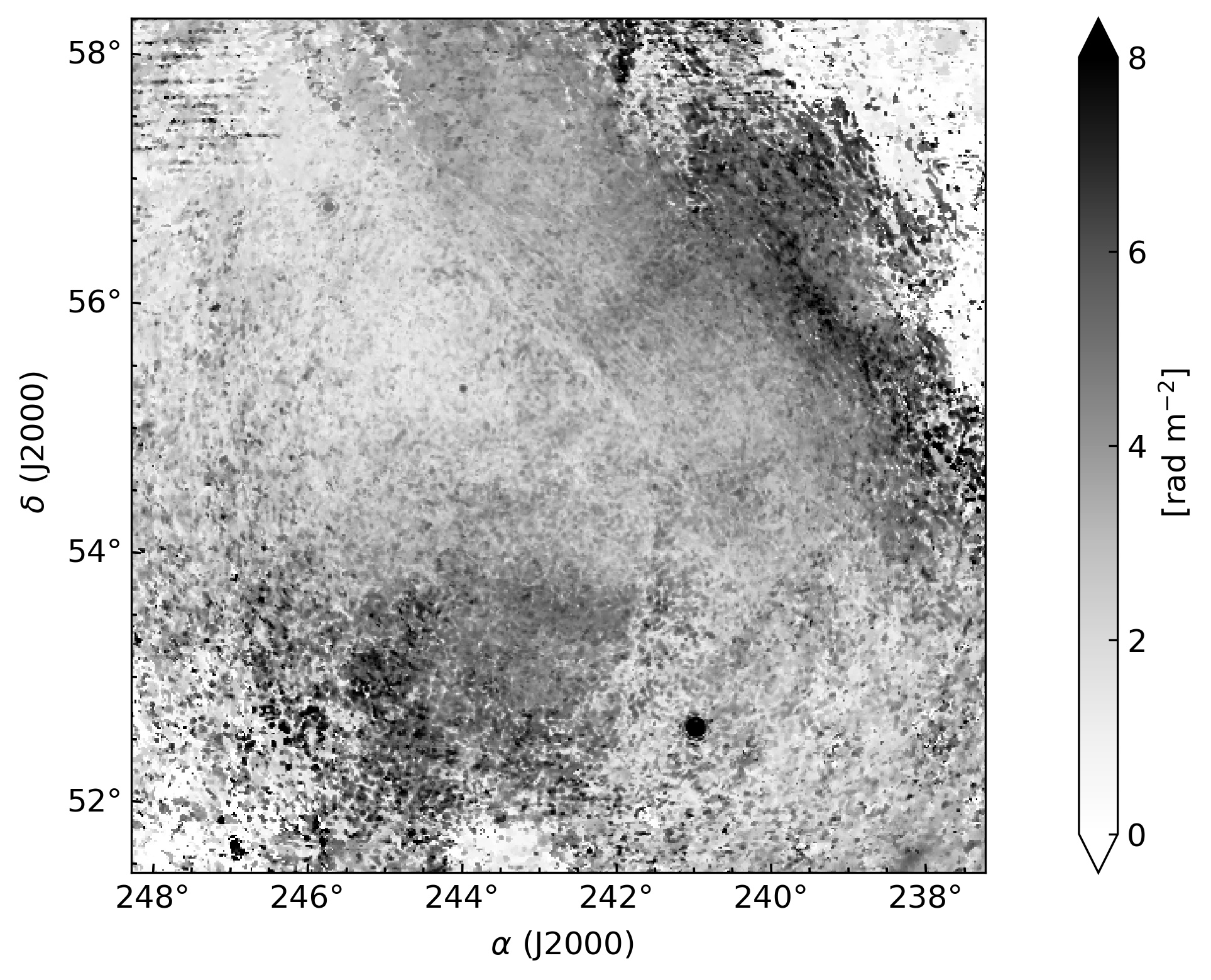}
  \caption{Moments of reference (014) (upper images) and final stacked Faraday cube (lower images). The left images give $M_0$, the middle ones show $M_1$, and the right images give $\sqrt{M_2}$.}
\label{fig:faraday_moments}
\end{figure*}

\subsection{Faint diffuse Galactic polarised emission}
\label{sub:diffuse}
We detect diffuse polarised emission in the final stacked Faraday cube over a range of Faraday depths from -16 up to +18 $\rm{rad\,m^{-2}}$ (see Appendix~\ref{app:supp}). Its brightest and prominent morphological features were already detected by \citet{jelic14}, but over a smaller Faraday depth range, ranging from -10 to +13 $\rm{rad\,m^{-2}}$, and with a poorer resolution of 1.75 $\rm{rad\,m^{-2}}$. Here we give a description of all morphological features observed in our final stacked cube.  

From $-16$ to $-4$ $\rm{rad\,m^{-2}}$ there is a northwest to the southeast gradient of emission. It starts as a small-scale feature in the northwest part of the image, and then it grows diagonally across the centre of the image to an extended northeast-southwest structure. Its mean surface brightness is 3.1 $\rm{\mu Jy\,PSF^{-1}\,RMSF^{-1}}$. From $-4$ to $-0.5$ $\rm{rad\,m^{-2}}$ there is diffuse emission whose morphology is more patchy, but it spreads over the full FoV. It has a mean surface brightness of 3.5 $\rm{\mu Jy\,PSF^{-1}\,RMSF^{-1}}$. A conspicuous, stripy morphological pattern of diffuse emission with north-to-south orientation dominates in the eastern part of the image from $+0.5$ up to $+4~\rm{rad\,m^{-2}}$. Its mean surface brightness reaches 4.3 $\rm{\mu Jy\,PSF^{-1}\,RMSF^{-1}}$. Towards higher Faraday depths, structures become very patchy, emission gets fainter, and then it disappears completely at +18 $\rm{rad\,m^{-2}}$. The mean surface brightness of this faint emission is 0.4 $\rm{\mu Jy\,PSF^{-1}\,RMSF^{-1}}$.

We constructed Faraday moments to make a comparison between the observed diffuse emission in the final stacked cube and the reference (014) observation. Faraday moments provide a statistical description of Faraday tomographic cubes, as introduced by \citet{dickey19}. The zeroth Faraday moment, $M_0$, is the polarised intensity $PI(\Phi)$ integrated over the full Faraday depth range, given in units of ${\rm mJy~PSF^{-1}~RMSF^{-1}~rad~m^{-2}}$. It gives the total polarised brightness of the emission in the Faraday cube. The first Faraday moment, $M_1$, is the polarised intensity weighted mean of Faraday spectra in units of ${\rm rad~m^{-2}}$.  It measures a mean Faraday depth at which the brightest emission is observed. Finally, the second Faraday moment, $M_2$, is the intensity-weighted variance of Faraday spectra, whose square root gives the spread of the spectrum in units of ${\rm rad~m^{-2}}$. Its square root measures a range of Faraday depths over which the brightest emission is observed. The Faraday moments are defined as
\begin{equation}
M_0=\sum_{i=1}^{n}PI_i\cdot\Delta\Phi,
\end{equation}
\begin{equation}
M_1=\frac{\sum_{i=1}^{n}PI_i\cdot\Phi_i}{\sum_{i=1}^{n}PI_i},
\end{equation}
and
\begin{equation}
M_2=\frac{\sum_{i=1}^{n}PI_i\cdot(\Phi_i-M_1)^2}{\sum_{i=1}^{n}PI_i},
\end{equation}
where $\Delta\Phi$ is a step in Faraday depth. The Faraday moments are calculated only for emission whose brightness is larger than a defined threshold to exclude noise-dominated areas in the data. We used a threshold of $m_P+5\sigma_P$, where $m_P$ is the polarised intensity bias and $\sigma_P$ is noise in the polarised intensity.

Figure~\ref{fig:faraday_moments} shows the calculated Faraday moments, both for the reference (upper images) and the final stacked Faraday cube (lower images). There is around 15\% more integrated emission in the stacked cube than in the reference observation, as measured by the $M_0$. This is due to a better signal-to-noise ratio in the stacked cube than the reference cube and the contribution of the detected faint emission to the $M_0$. The first Faraday moments do not differ much, as they are mostly driven by the brightest emission, which is the same in both cases. However, mean values of the $M_1$ are $-0.85~{\rm rad~m^{-2}}$ and $-0.65~{\rm rad~m^{-2}}$ for reference and stacked cubes, respectively, indicating that there is on average more emission at positive Faraday depths in the stacked cube than in the reference cube. The $\sqrt{M_2}$ shows the most noticeable differences. The measured spread in Faraday depth is on average 42\% larger in the stacked than the reference cube. This is again due to faint emission at larger Faraday depths, which does not contribute to the second moment of the reference observation.

\begin{figure*}[t!]
\centering
  \includegraphics[width=0.3\linewidth]{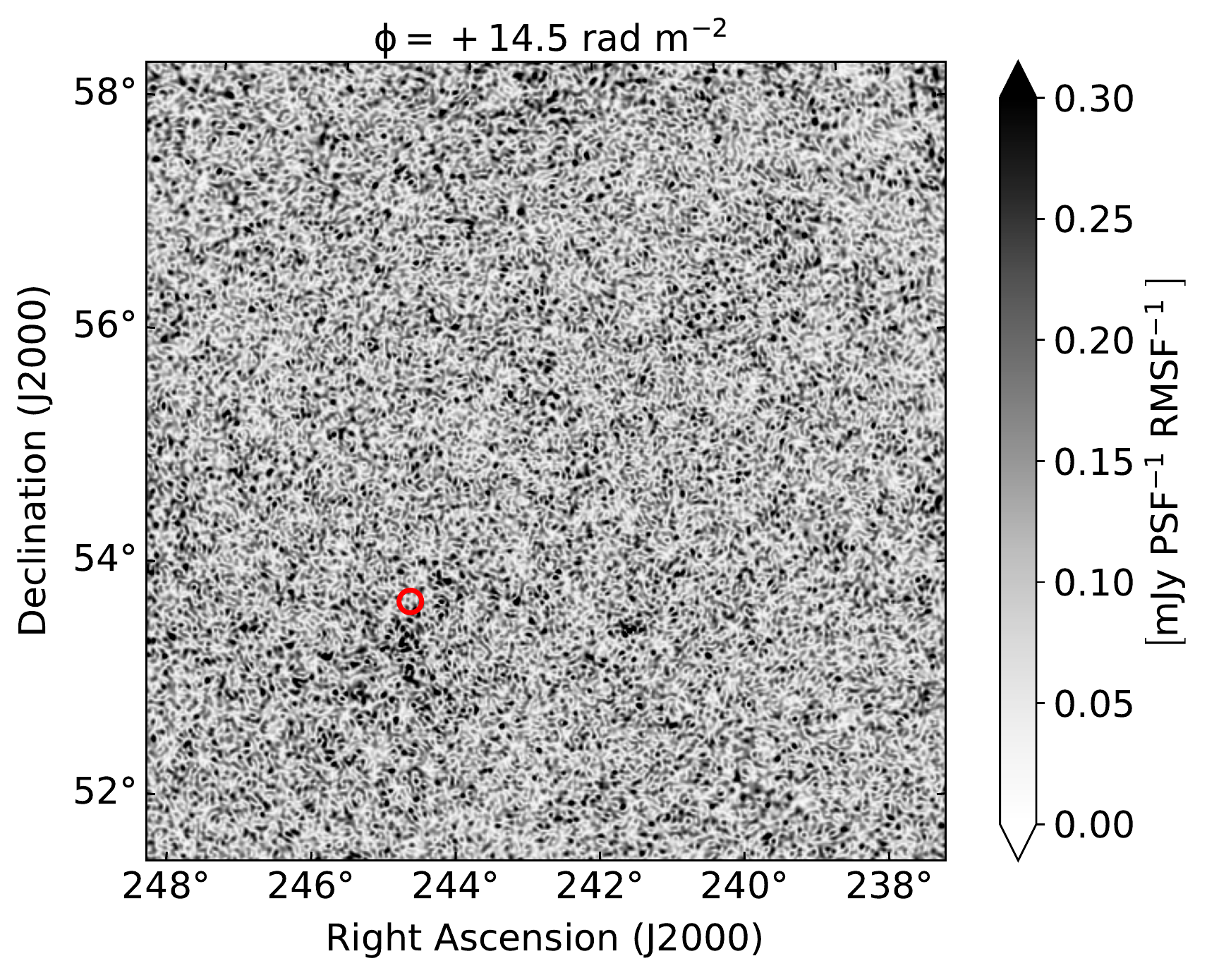}
  \includegraphics[width=0.3\linewidth]{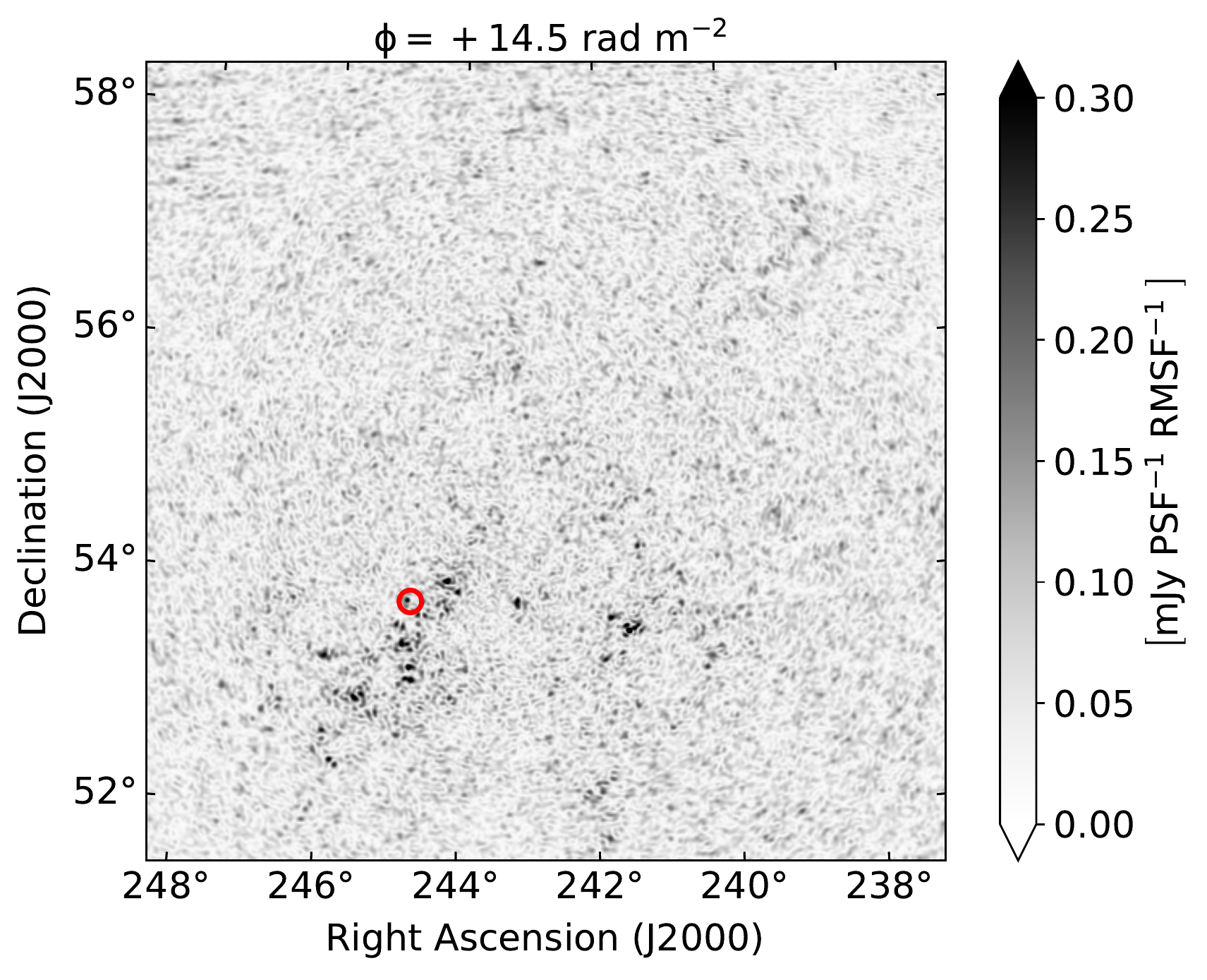}
  \includegraphics[width=0.33\linewidth]{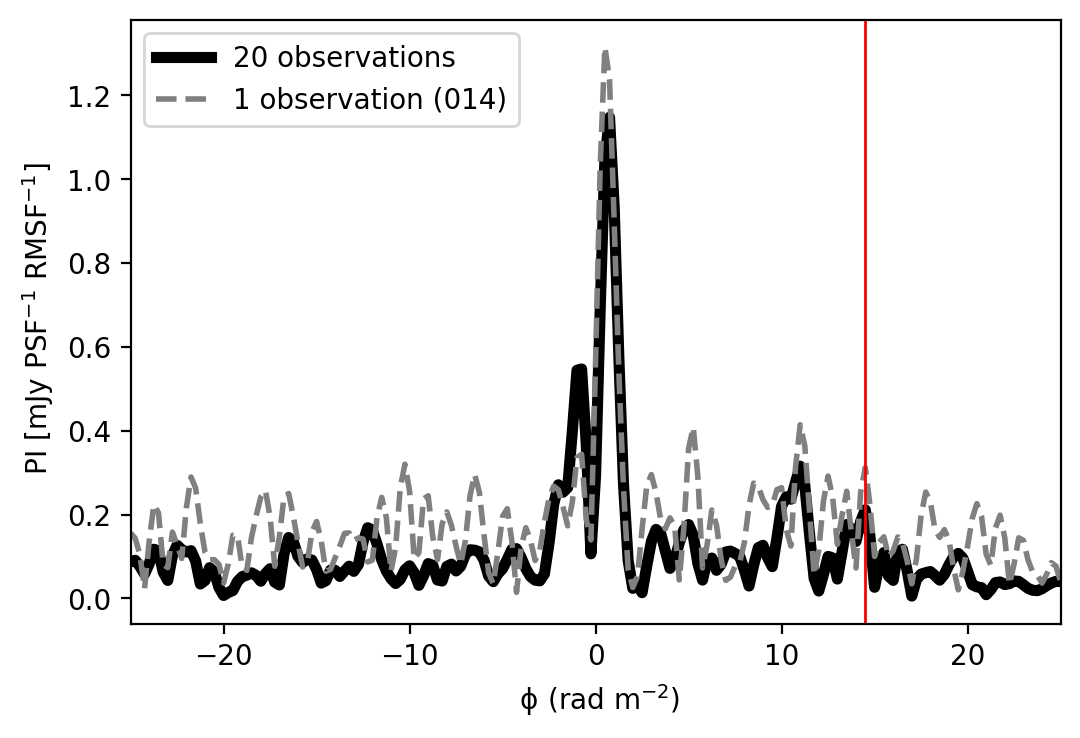}
  \includegraphics[width=0.3\linewidth]{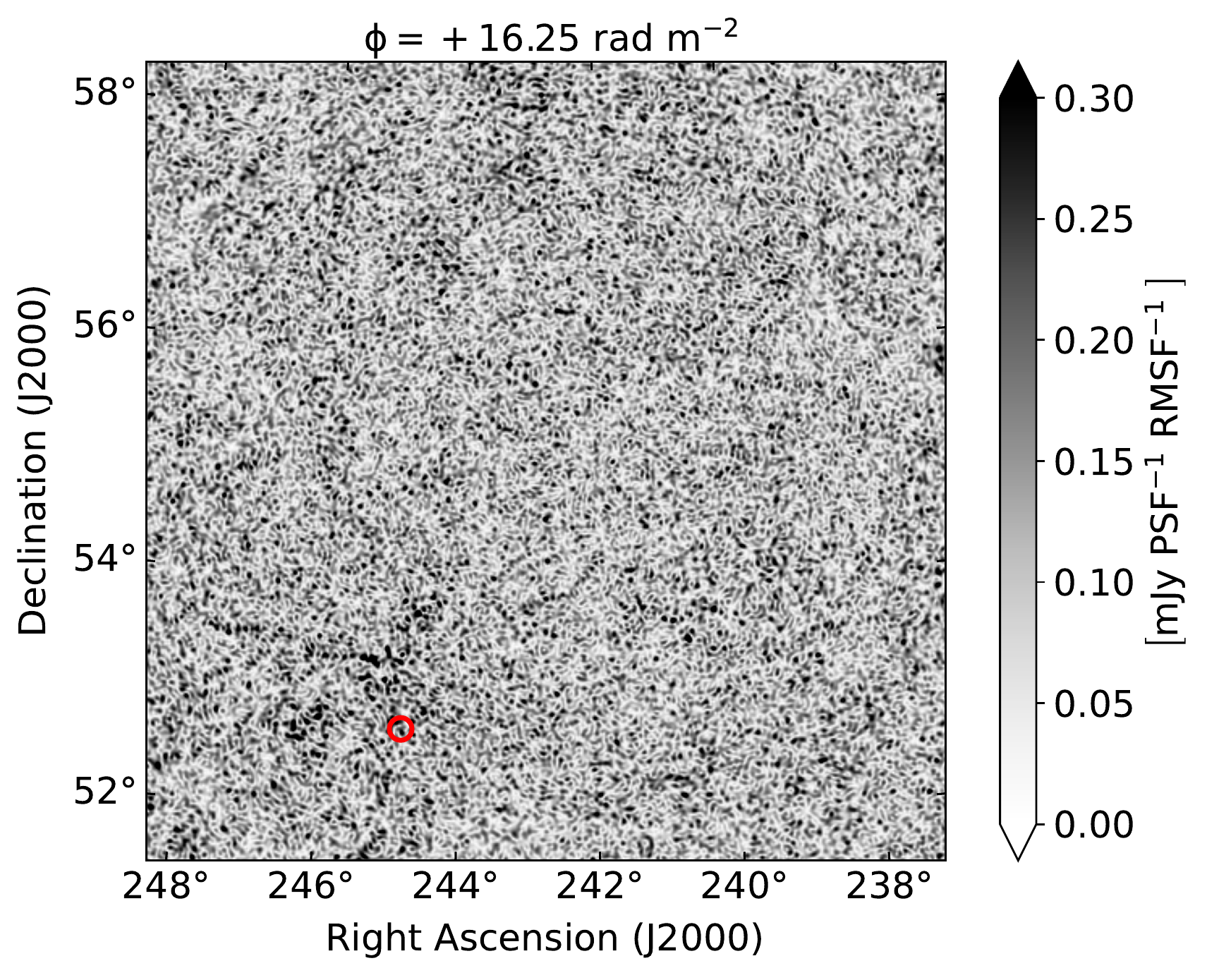}
   \includegraphics[width=0.3\linewidth]{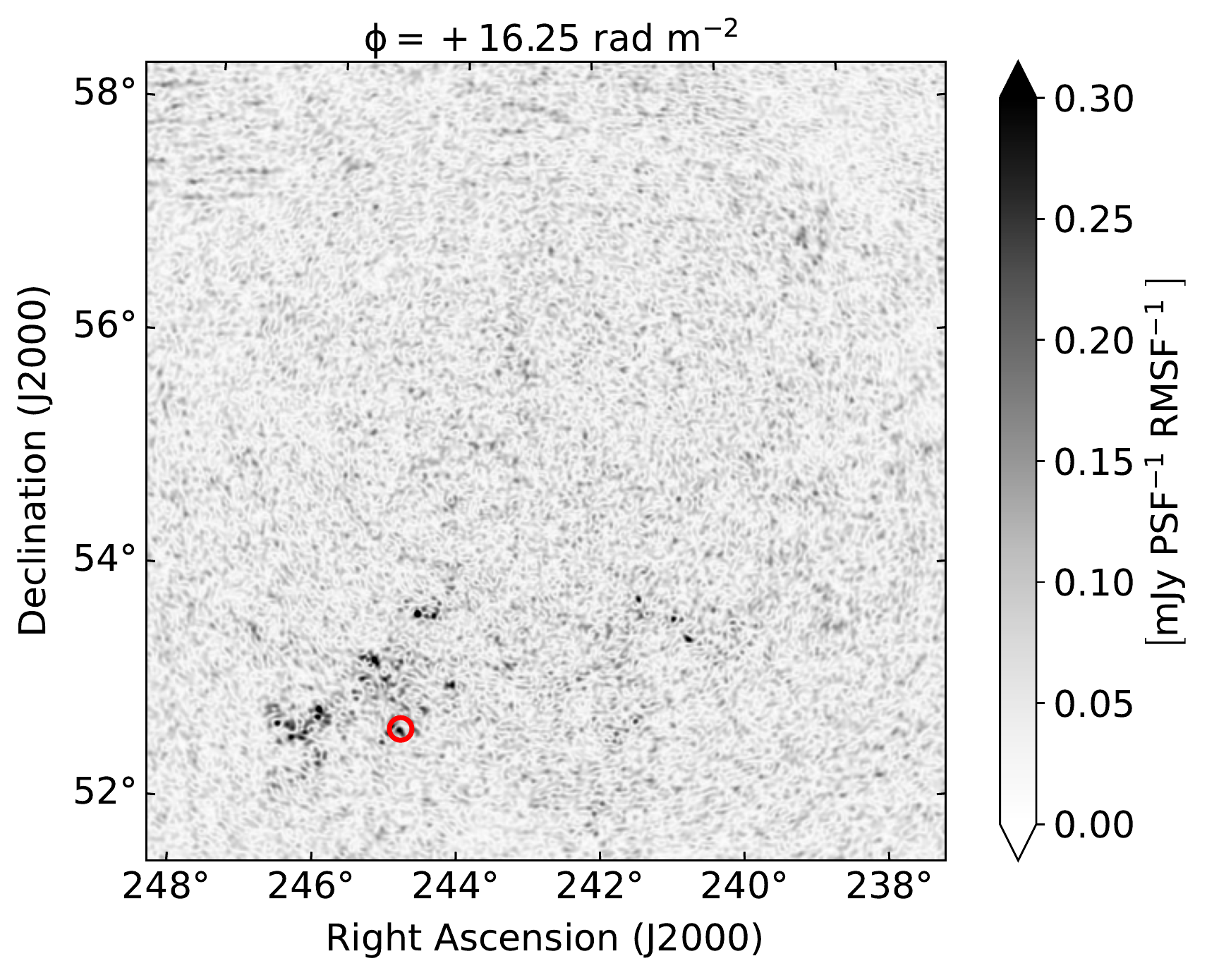}
  \includegraphics[width=0.33\linewidth]{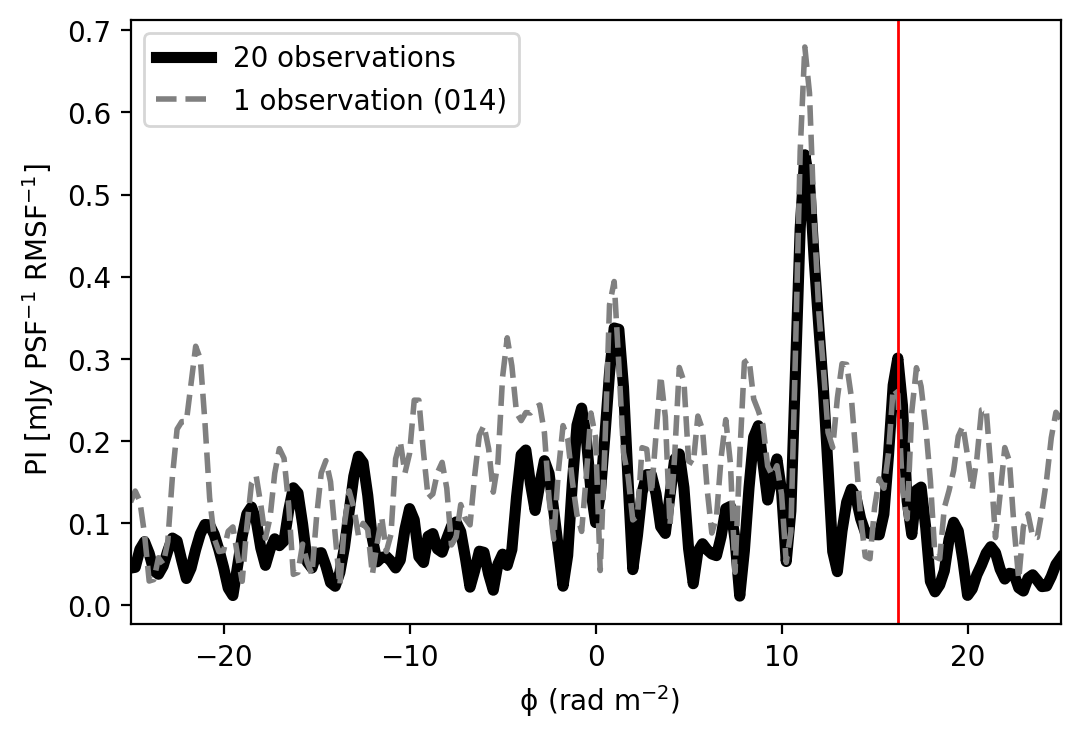}
\caption{Examples of faint Galactic polarised emission, which is only clearly detected in the final stacked Faraday cube. The images are given in the polarised intensity at Faraday depths of $+14.5$ (upper images) and $+16.25~{\rm rad~m^{-2}}$ (lower images) for the reference (left images) and the final stacked cube (middle images). The corresponding Faraday spectra at a location of the red circle in the images are given in plots on the right.}
\label{fig:diffuse_emission_examples}
\end{figure*}

Examples of the faint emission, which is only clearly detected in the stacked cube, are shown in Fig.~\ref{fig:diffuse_emission_examples}. The images are given at Faraday depths of $+14.5$ (upper images) and $+16.25~{\rm rad~m^{-2}}$ (lower images). The images in the first panels are shown for the reference cube, while in the second panels for the final stacked cube. The third panels show the corresponding Faraday spectra at a location of the red circle in the images. The brightness of the faint emission is comparable to the noise in the reference cube and therefore is not detected there.

\section{Discussion on the faint polarised emission newly detected}
The diffuse polarised emission detected in the final stacked Faraday cube has the mean polarised intensity of $10~{\rm mJy~PSF^{-1}~RMSF^{-1}~rad~m^{-2}}$, as measured in the central region of the $M_0$ (see bottom left image in Fig.~\ref{fig:faraday_moments}). This translates to a mean brightness temperature of $\sim9.5~{\rm K}$\footnote{The intensity of $1~{\rm mJy~PSF^{-1}~RMSF^{-1}~rad~m^{-2}}$ corresponds to a brightness temperature of $\sim0.95~{\rm K}$ at $144~{\rm MHz}$, a frequency that corresponds to the weighted average of the observed $\lambda^2$ used in RM synthesis.}.

We recalculate the $M_0$ of the final stacked Faraday cube by restricting it to Faraday depths  $\geq+13~{\rm rad~m^{-2}}$ to estimate the mean brightness temperature of the newly detected faint emission at higher Faraday depths. We get its mean polarised intensity to be of $0.5~{\rm mJy~PSF^{-1}~RMSF^{-1}~rad~m^{-2}}$, which is $\sim 0.475~{\rm K}$. Although this faint emission is not contributing more than $\sim 5~\%$ to the total observed polarised emission, its relevance comes from the fact that it is present at Faraday depths at which the emission was not observed before. It increases the range of Faraday depths, usually characterised by $\sqrt{M_2}$, over which the emission is detected in this field with LOFAR. This is especially important for the interpretation of the LOFAR observations regarding an extent of the probed volume along the LOS and underlying distribution of synchrotron-emitting and Faraday-rotating regions. 

Depolarisation effects associated with Faraday rotation are significant at low radio frequencies. Only a few percent of the intrinsically polarised synchrotron emission is observed with the LOFAR \citep{jelic14, jelic15, vaneck17, vaneck19, turic21}. The questions that arise relate to where along the LOS does depolarisation happen and from where does the observed emission originate. The idea is that we observe mostly close-by emission, while far-away emission is depolarised in the magneto-ionic medium on the way to us. However, determining this from the LOFAR observations alone is very difficult. A Faraday depth is not necessarily a good proxy for the distance.  We need to take into consideration the full-complexity of the magnetic fields, its possible reversals, and the multi-phase nature of the interstellar medium. 
This is challenging, but it has been attempted recently in a number of the multi-tracer and -frequency studies of the LOFAR observations \citep[][]{zaroubi15,vaneck17,jelic18,bracco20,turic21,erceg22} and by using the magneto-hydrodynamical simulations \citep[][]{bracco22}. For example, \citet{erceg22} compared the Faraday moments of the LOFAR observations of around 3100 square degrees in the high-latitude outer Galaxy to the high-frequency polarisation data \citep[DRAO GMIMS,][]{dickey19} and to the Galactic Faraday Sky map \citep{hutschenreuter22}. The latter compliments the low- and high-frequency observations, as it represents the total RM yielded from the Galaxy. It is constructed using the observed RM of a large sample of extragalactic polarised sources, including the one in the LoTSS polarised source catalogue \citep{OSullivan23}. 

\citet{erceg22} found a correlation between the Galactic Faraday Sky map and the LOFAR first Faraday moment image. However, the ratio of the two cannot be explained by a simple model of a Burn slab \citep{burn66}, which seems to be applicable to the high-frequency data \citep{ordog19}. A Burn slab assumes a mixture of uniform synchrotron-emitting and Faraday-rotating regions along the LOS and predicts a ratio of two between the modelled total Galactic RM and the observed polarised emission. The observed LOFAR Faraday spectra are more complex to understand, highlighting the high level of complexity of the LOS distribution of synchrotron emission and Faraday rotation.

\begin{figure}[t!]
    \centering    \includegraphics[width=\linewidth]{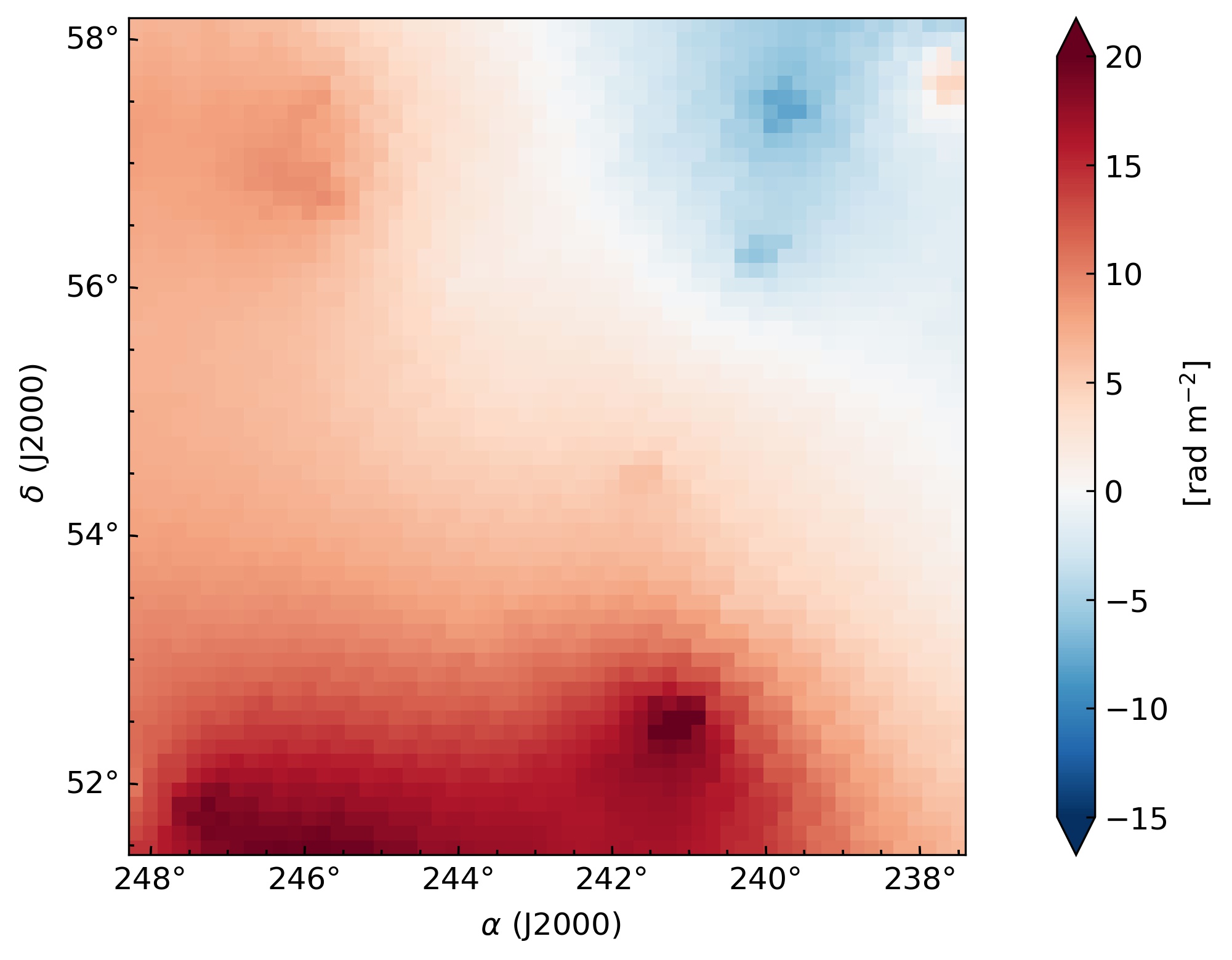}
    \caption{Total Galactic RM  in the area of the ELAIS-N1 field, extracted from the publicly available the Galactic Faraday Sky map by \citet{hutschenreuter22}.}
    \label{fig:hutschen}
\end{figure}

We compare the $M_1$ of our stacked Faraday cube with the Galactic Faraday Sky map \citep{hutschenreuter22}. Figure~\ref{fig:hutschen} shows a cut-out of this map in the area of ELAIS-N1 field. A visual comparison with the bottom middle panel of Fig.~\ref{fig:faraday_moments} shows that the northwest-to-southeast gradient in the first moment is also present in the Galactic Faraday sky map. The values are more negative around the northwest corner of the image; then, diagonally towards the centre of the image they verge towards zero, and then they increase to more positive values towards the southeast corner of the image. This gradient implies a bending magnetic field in a southeast to a northwest direction. The magnetic field mostly points towards us in the southeast corner of the image, in the central part of the image it is mostly in the plane of the sky, and then in the northwest corner it points away from us. 

Comparable negative values towards the northwest corner of the image between the two maps means that we are probing the same volume of magneto-ionic medium. On the contrary, larger positive values of Galactic Faraday Sky towards the southeast corner of the image than in the first moment means that we are not probing the same volume. Emission that comes from further away is either depolarised in LOFAR observations or it is Faraday thick, as discussed by \citet{erceg22} in a broader discussion of Faraday moments in the high-latitude outer Galaxy. Moreover, from the same work \citep[see Figs 5, 8, 9, and 10 in][]{erceg22}, it is clear that the ELAIS-N1 field is just at the edge of a region associated with the polarised emission from the radio Loop III. The observed gradient is perpendicular to the shape of the loop, and it is probably associated with it. The same is true for the emission observed in the surrounding area of the ELAIS-N1 field. 

Faraday depths at which we observe the diffuse emission in the ELAIS-N1 field are comparable or smaller than the total Galactic RM in the same area. Therefore, the observed diffuse emission is probably Galactic, including its newly detected faint component.

\section{Summary and conclusions}
\label{sub:summ}
We used 21 LOFAR HBA observations of the ELAIS-N1 field (about 150 hours of data) to conduct the deepest polarimetric study of Galactic synchrotron emission at low radio frequencies to date. The analysis was performed on very low-resolution (4.3') Stokes QU data cubes produced as part of the LoTSS-Deep Fields Data Release 1 \citep{sabater21}. A stacking technique was developed to improve sensitivity of the data based on diffuse polarised emission. The outcomes of this analysis are the following: 

\begin{enumerate}
    \item We verified the reliability of the absolute ionospheric Faraday rotation corrections estimated using the satellite-based TEC measurements to be of $\sim0.05~{\rm rad~m^{-2}}$. We also demonstrated that diffuse polarised emission itself can be used to account for the relative ionospheric corrections with respect to some reference observation.
    \item We showed the feasibility of the developed stacking technique by combining 20 single-night observations into one 150 hour data set. The resulting Faraday cube has a noise of $27~{\rm \mu Jy~PSF^{-1}~RMSF^{-1}}$ in polarised intensity, which is an improvement of $\sim\sqrt{20}$ in comparison with noise in an individual five-to-eight-hour observation. 
    \item The rotation measures of successfully detected polarised sources in our final Faraday cube are in agreement with the values provided in the catalogue by \citet{ruiz21}, which is based on the higher angular resolution LoTSS Deep Field images.
    \item We detected a faint component of diffuse polarised emission in the stacked cube at high Faraday depths, ranging from $+13$ to $+17~{\rm rad~m^{-2}}$, which was not detected in the commissioning observation by \citet{jelic14}. The brightness temperature of this emission is $~\sim475~{\rm mK}$, which is almost an order of magnitude fainter than the brightest emission observed in this field.
    \item The observed northwest to southeast gradient of emission in Faraday depth  we associate with a bending magnetic field across the FoV. It is probably connected to the radio Loop III, as the ELAIS-N1 field is just at the edge of it. 
\end{enumerate}
The presented stacking technique provides a valuable tool and gives perspective for the future deep polarimetric studies of Galactic synchrotron emission at low radio frequencies with LOFAR,  the Square Kilometre Array, and its other precursors. For example, it can be applied to other LoTSS Deep Fields (Lockman Hole and Bo\"otes), as well as to the Great Observatories Origins Deep Survey - North (GOODS-N) field. It is also of importance for cosmological studies, where the polarised emission of the Milky Way is the foreground contaminant. For instance, if one wants to measure magnetic field properties in the cosmic web \citep[e.g.][]{carretti22}, it is crucial to have an independent measurement of the Galactic contribution to the total Faraday rotation observed toward extragalactic sources. Moreover, successful extraction of the cosmological signal from the cosmic dawn and epoch of reionisation also relies on good knowledge of the foreground emission, including the Galactic polarised emission \citep[e.g.][]{jelic10, asad15,spinelli19}.

\begin{acknowledgements}
We thank the anonymous referee for their constructive comments. VJ, AE, LC and LT acknowledge support by the Croatian Science Foundation for a project IP-2018-01-2889 (LowFreqCRO) and additionally LT and VJ for the project DOK-2018-09-9169. PNB is grateful for support from the UK STFC via grant ST/V000594/1. MH acknowledges funding from the European Research Council (ERC) under the European Union's Horizon 2020 research and innovation program (grant agreement No 772663). VV acknowledges support from Istituto Nazionale di Astrofisica (INAF) mainstream project “Galaxy Clusters Science with LOFAR”, 1.05.01.86.05. This paper is based on data obtained with the International LOFAR Telescope (ILT) under project codes LC0\_019, LC2\_024 and LC4\_008. LOFAR (van Haarlem et al. 2013) is the Low Frequency Array designed and constructed by ASTRON. It has observing, data processing, and data storage facilities in several countries, that are owned by various parties (each with their own funding sources), and that are collectively operated by the ILT foundation under a joint scientific policy. The ILT resources have benefited from the following recent major funding sources: CNRS-INSU, Observatoire de Paris and Université d’Orléans, France; BMBF, MIWF-NRW, MPG, Germany; Science Foundation Ireland (SFI), Department of Business, Enterprise and Innovation (DBEI), Ireland; NWO, The Netherlands; The Science and Technology Facilities Council, UK; Ministry of Science and Higher Education, Poland.
\end{acknowledgements}

%
%

\bibliographystyle{aa} 
\bibliography{bibliography_1.bib}
\appendix 
\twocolumn
\section{Restoring the 011 observation using Galactic polarised emission}
\label{app:restoring}

\begin{figure*}[h!]
\includegraphics[width=0.33\linewidth]{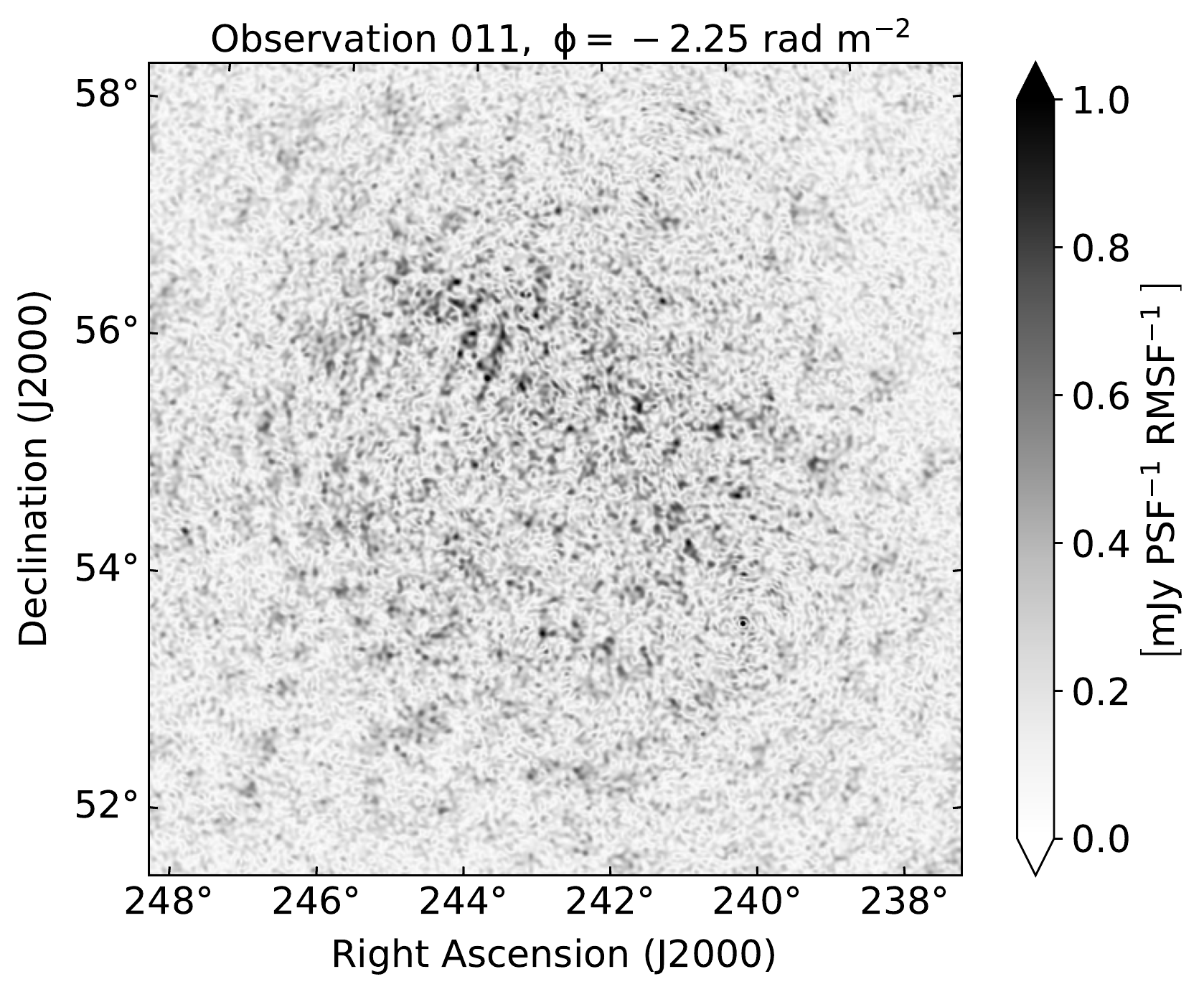}
\includegraphics[width=0.33\linewidth]{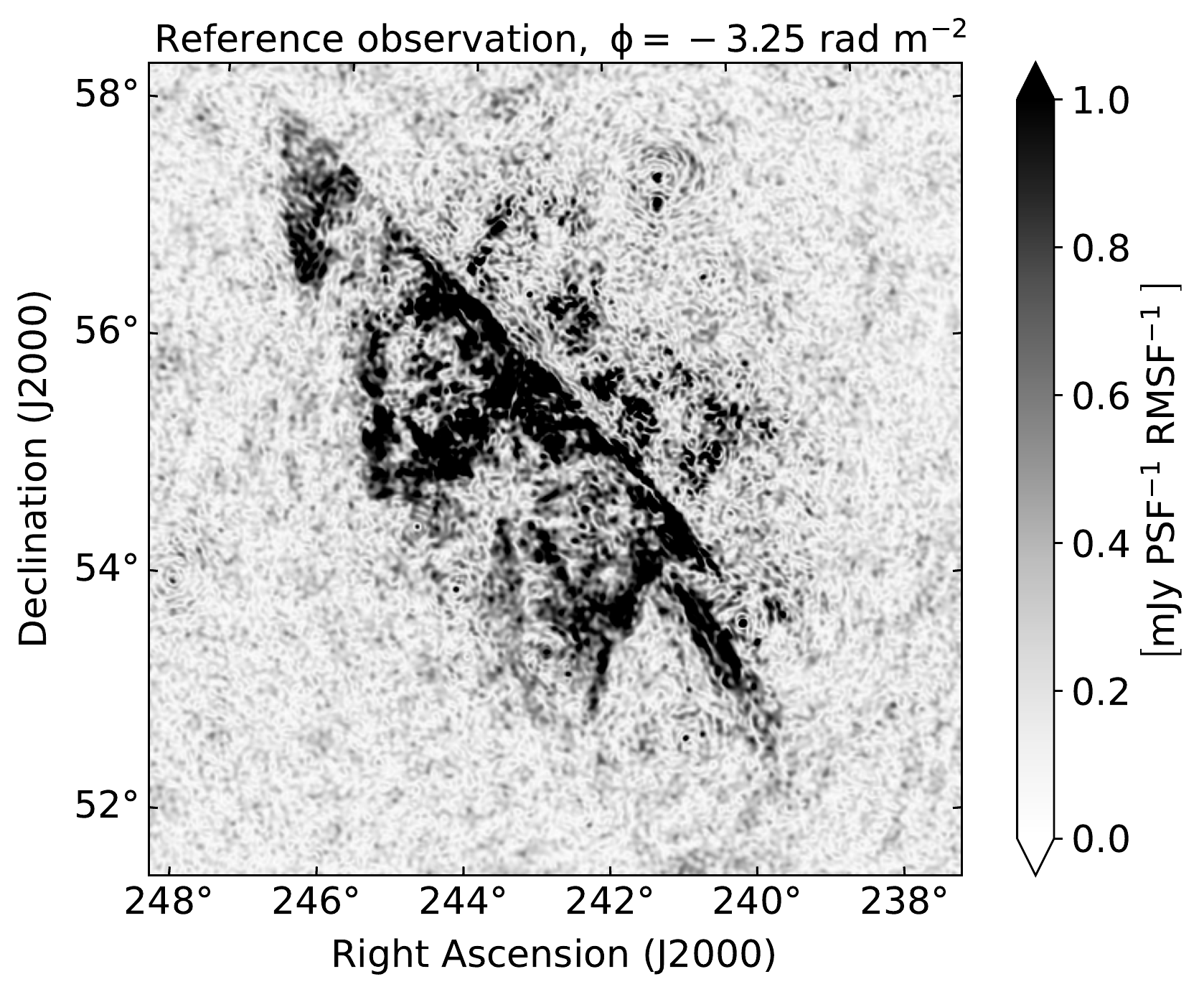}
\includegraphics[width=0.33\linewidth]{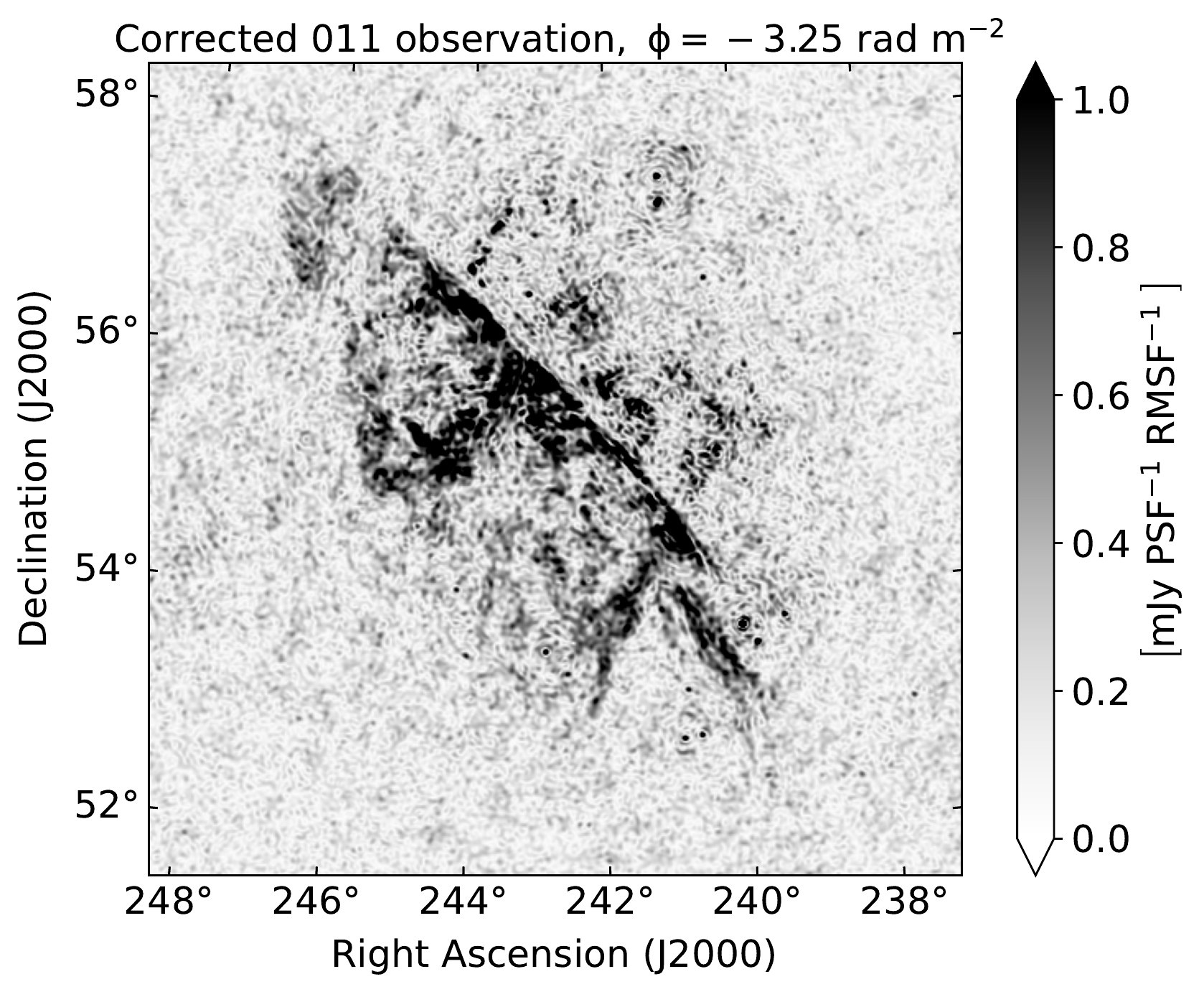}
\caption{Example of image in Faraday cube given in the polarised intensity at $\rm{-2.25\,rad\,m^{-2}}$ for the 011 observation (left image), which is not properly corrected for the $RM_{\rm ion}$. There is almost no emission visible in comparison to the reference (014) observation (middle image), whose image is given at $\rm{-3.25\,rad\,m^{-2}}$  to account for a relative misalignment of $\rm{+1.0\,rad\,m^{-2}}$ between the two observations. The polarised emission is visible in the restored Faraday cube of the 011 observation (right image), which is corrected using the estimated $\Delta\Phi_{\rm shift}$ given in Fig.~\ref{fig:10minutes_change}.}
\label{fig:image_corrected}
\end{figure*}

The 011 observation shows the largest shift with respect to the reference observation ($\rm{\gtrsim1\,rad\, m^{-2}}$) among all observations analysed in this work. We inspected the Faraday cube for this observation in the polarised intensity. There is almost no emission visible in the Faraday cube in comparison to the reference observation. An example is given in Fig.~\ref{fig:image_corrected} (left image) at a Faraday depth of $\rm{-2.25\,rad\,m^{-2}}$. The same figure (middle image) shows an image of the reference observation, but at a Faraday depth of $\rm{-3.25\,rad\,m^{-2}}$ to account for a relative misalignment between the two observations in Faraday depth. The lack of the observed polarised emission shows that $RM_\mathrm{ion}$ corrections were not applied properly to the data due to some unfortunate processing error. We confirm this by inspection of the processing log files. 

The calculated $RM_\mathrm{ion}$ correction for this observation is $2.4~{\rm rad~m^{-2}}$ at the beginning of the observation, and then it decreases to $1.5~{\rm rad~m^{-2}}$ within the first 430 minutes. This relative change of $0.9~{\rm rad~m^{-2}}$ is enough to fully depolarize the signal at 150 MHz (see Sect.~\ref{sec:intro}), if we do not correct the data for it. In the remaining 50 minutes of the observation, it increases again to $1.6~{\rm rad~m^{-2}}$. 

To restore the polarised signal in 011 observation, we test if the observed polarised emission itself can be used to account for the ionospheric Faraday rotation correction that should be applied to the data. We first re-imaged the eight-hour 011 observation by creating 48 Stokes QU images of ten-minute intervals of the observation. Then, for each ten-minute interval, we found its relative shift in Faraday depth with respect to the full eight-hour reference observation following the methodology described in Sect.~\ref{sub:ionoEFF}. 

\begin{figure}[t!]
    \centering
    \includegraphics[width=\linewidth]{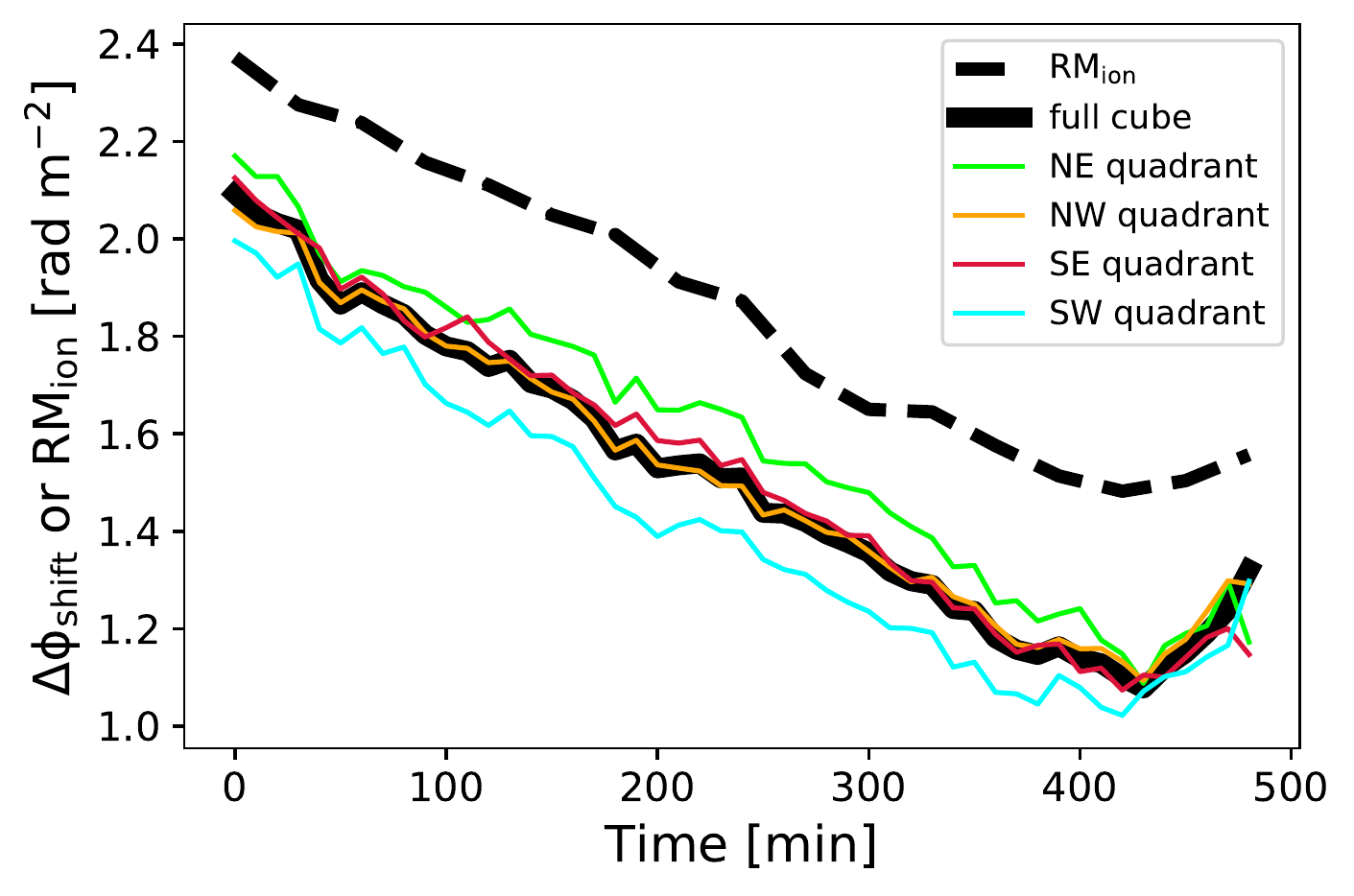}
     \caption{Estimated relative shifts in Faraday depth ($\Delta\Phi_{\rm shift}$) of each ten-minute interval of the 011 observation with respect to the full eight-hour reference (014) observation (thick solid black line). The calculated $RM_{\rm ion}$ corrections based on the satellite TEC measurements are plotted with a thick dashed black line. The thin solid coloured lines give the $\Delta\Phi_{\rm shift}$ in the field-of-view quadrants.}
       \label{fig:10minutes_change}
\end{figure}

Figure~\ref{fig:10minutes_change} shows the results (thick solid black line), which are compared with the corrections calculated using the satellite-based TEC measurements (black dashed line). The two curves are showing the same trend. A systematic shift of $\sim0.3~{\rm rad~m^{-2}}$ between the two curves is due to the different nature of these two methods. The satellite-based corrections give absolute $RM_\mathrm{ion}$, while the corrections based on the observed polarised emission give relative values with respect to the used reference observation ($\Delta\Phi_{\rm shift}$). 

We also tested for any angular variations of $\Delta\Phi_\mathrm{shift}$ across the FoV. We did this by splitting the frequency cube spatially into quadrants and then repeating the procedure to find a relative time varying shift for each quadrant separately. The results are over-plotted in Fig.~\ref{fig:10minutes_change} with thin solid coloured lines. The northwest (NW) and the southeast (SE) quadrants show  relative shifts which are consistent with the result of the full cube. Larger values are found in the northeast (NE) quadrant; on average, the shifts are larger by $0.096~{\rm rad~m^{-2}}$ than the one from the full cube. In the southwest (SW) quadrant, they are smaller by $0.1~{\rm rad~m^{-2}}$. This points to a relative spatial gradient of $\sim 0.2~{\rm rad~m^{-2}}$ in the northeast-southwest direction across the FoV.

We `de-rotated' the observed polarisation angle of each ten-minute interval by its estimated shift across the full image with respect to the reference observation ($\Delta\Phi_{\rm shift}$). This is done by multiplying the complex polarisation given at each wavelength (frequency) by $\exp^{-i2\Delta\Phi_{\rm shift} \lambda^{2}}$. Then, we combined all corrected ten-minute intervals to obtain a Stokes QU cube over the full eight-hour synthesis and used the RM synthesis to obtain the final restored Faraday cubes of the 011 observation. 

The polarised emission is now visible in the restored Faraday cubes. An example of polarised intensity is shown in Fig.~\ref{fig:image_corrected} (right image) at Faraday depth of $\rm{-3.25\,rad\,m^{-2}}$. Observed morphology of polarised emission in the restored 011 observation and the reference observation are visually very similar.  To quantify this similarity, we calculate the Pearson correlation coefficient between the images of the highest peak value of the Faraday depth spectrum in the polarised intensity of the two observations and a ratio of their peak intensity distributions. We are using in the calculation only the inner $3^\circ\times3^\circ$ of the images. We get a correlation coefficient of $0.95$, and find that polarised emission in the restored cube is $(73\pm14)\%$ of that in the reference observation. The majority of the emission and its morphology is restored.

To increase the percentage of the recovered brightness of the observed emission even further, we would need to address the depolarisation that happens within ten-minute intervals. To achieve that, we could re-image the eight-hour observation to even smaller time intervals, for example of one minute. However, the signal-to-noise ratio in that case would be a limiting factor for our methodology, making it out of the scope of current work. 

\begin{figure}
    \centering
    \includegraphics[width=\linewidth]{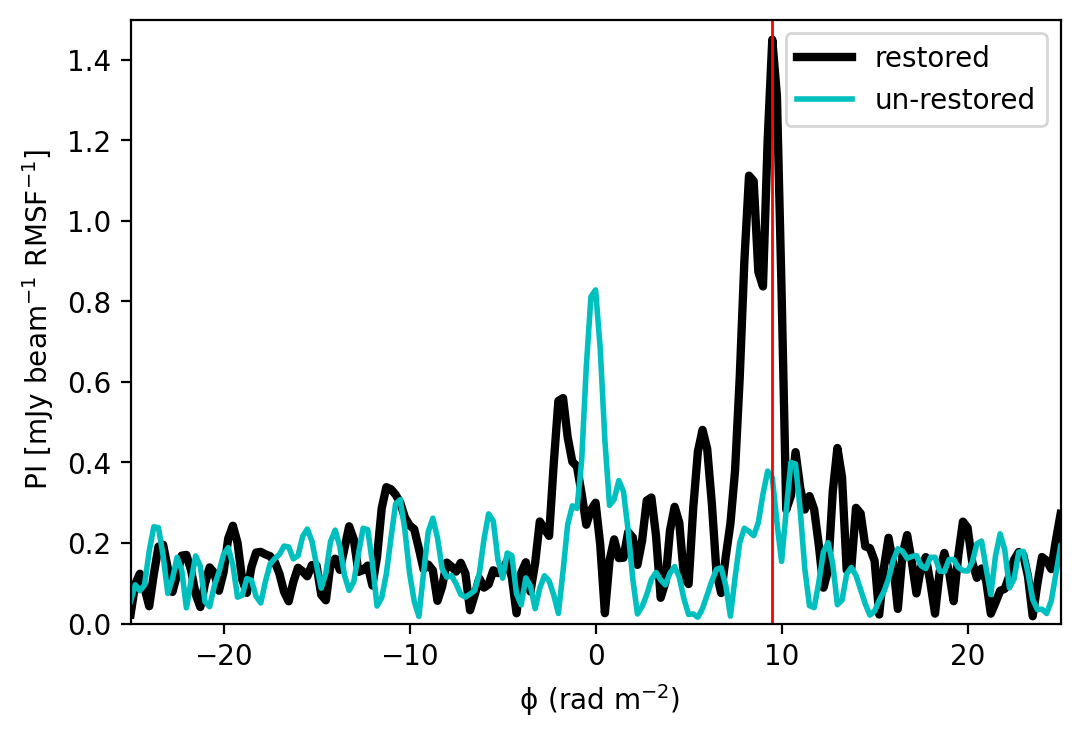}
    \caption{Primary-beam-uncorrected Faraday spectrum at a location of a polarised source \citep[ID 01 in][]{ruiz21} in the high-resolution restored (solid black line) and unrestored (solid cyan line) Faraday cube of the 011 observation. The source should be present at a Faraday depth of $\rm{+(9.44\pm0.03)\,rad\,m^{-2}}$  (red vertical line), as reported by \citet{ruiz21}.}
    \label{fig:Hresolution}
\end{figure}

Furthermore, we assess if the estimated shifts can be applied to the high-resolution images ($6''$) of the same observation.  We re-imaged a small part of the high-resolution data centred at a polarised source at RA $16^{\rm h}24^{\rm m}32^{\rm s}$ and Dec $+56^{\circ}52'28''$ \citep[][source with ID 01]{ruiz21} in ten-minute intervals.  Then, we `de-rotated' the observed polarisation angle of each ten-minute interval by the shift estimated using the very-low resolution data, combined high-resolution ten-minute intervals with the full eight-hour synthesis frequency cube, and applied the RM synthesis to it. The resulting primary-beam-uncorrected Faraday spectrum at a location of the polarised source is presented in Fig.~\ref{fig:Hresolution} (black line). The source appears at a Faraday depth of $\rm{\sim9.5\,rad\,m^{-2}}$, as expected.  Its recovered peak polarised flux is $43\%$ of the value reported by \citet{ruiz21}, once we take into account the primary beam correction at the location of the source (a factor of $2.3\times$). The same figure also shows the Faraday spectrum before the ionospheric corrections are applied (cyan line), where the source is fully depolarised. A successful detection of the source demonstrates a potential of using the very low-resolution data to correct the high-resolution data. This method is computationally more efficient than the one  that uses the high-resolution data only.

\newpage
\onecolumn

\section{Supplementary figures}\label{app:supp}

This appendix presents images in the polarised intensity of the final stacked Faraday cube described in Sect.~\ref{sub:diffuse}. 
\begin{figure*}[!h]
\begin{center}
\resizebox{1.5\hsize}{!}{\includegraphics{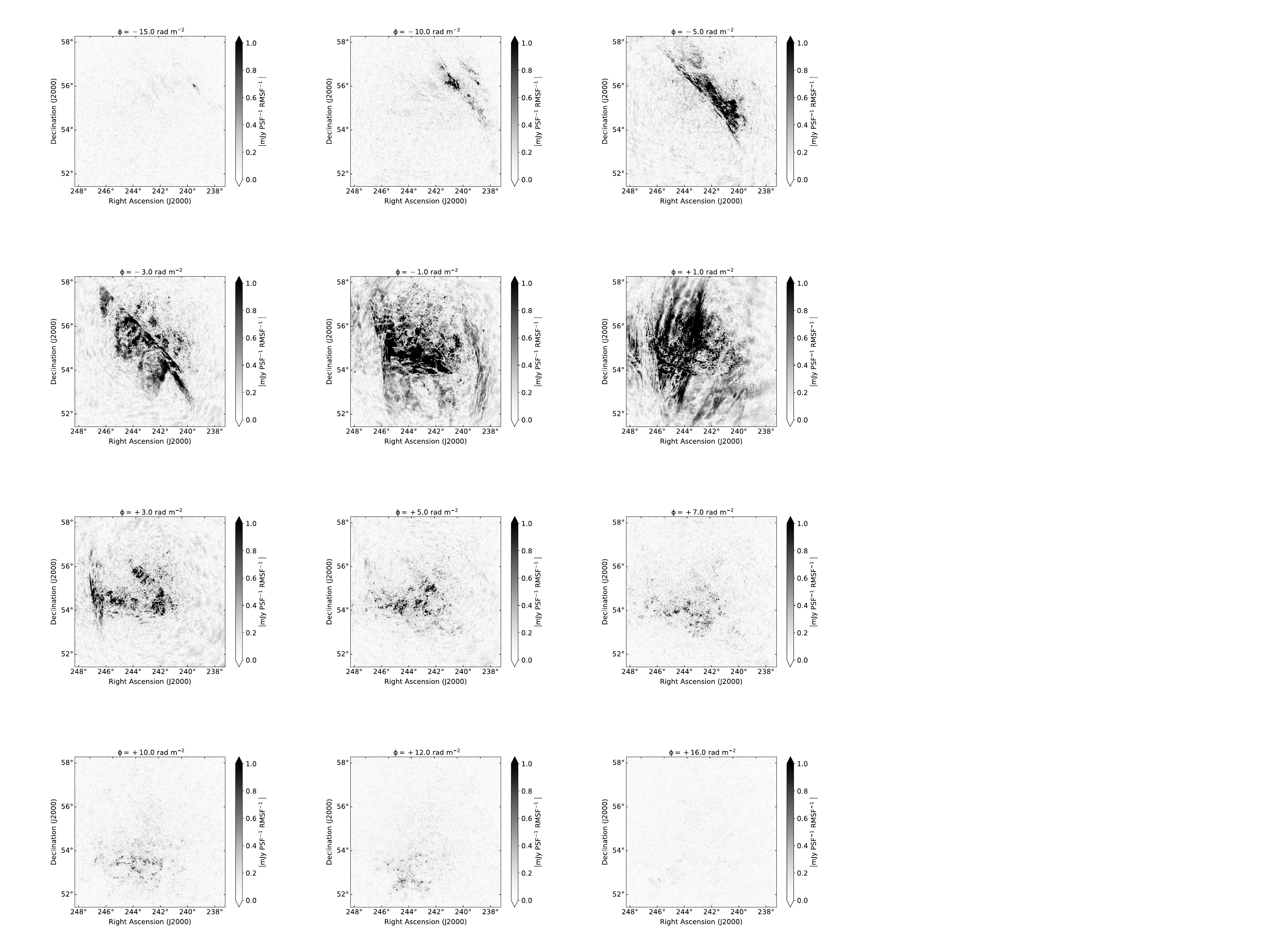}}
\caption{Images of the ELAIS-N1 field in the polarised intensity given at Faraday depths of -15.0, -10.0, -5.0, -3.0, -1.0, +1.0, +3.0, +5.0, +7.0, +10.0, +12.0 and +16.0 $\rm{rad\,m^{-2}}$  of the final stacked Faraday cube. The cube is based on $\sim150$ hours of the LOFAR observations in the frequency range from 114.9 to 177.4 MHz. Angular resolution of the images is $4.3'$. These are primary-beam-uncorrected images with the noise of $27~{\rm \mu Jy~PSF^{-1}~RMSF^{-1}}$.}
\label{fig:stacked_cube}
\end{center}
\end{figure*}

\end{document}